\documentclass[a4paper,12pt]{article}
\pdfoutput=1
\pdfoutput=1
\usepackage{enumerate}
\usepackage{jheppub,undertilde}
\usepackage{amsthm}
\usepackage{graphicx}
\usepackage{color}
\usepackage{siunitx}
\usepackage{siunitx}
\usepackage{graphicx,textcomp,float,gensymb,wrapfig, enumitem,comment,dsfont,subfigure,framed,slashed,appendix,wrapfig}
 \usepackage[colorinlistoftodos]{todonotes}
 \usepackage{hyperref}

 \newcommand{\be}{\begin{equation}}
\newcommand{\ee}{\end{equation}}
\def\bsp#1\esp{\begin{split}#1\end{split}}

\newcommand{\bea}{\begin{eqnarray}}  
\newcommand{\eea}{\end{eqnarray}}  

\newcommand{\appropto}{\mathrel{\vcenter{
  \offinterlineskip\halign{\hfil$##$\cr
    \propto\cr\noalign{\kern2pt}\sim\cr\noalign{\kern-2pt}}}}}

\preprint{SISSA 29/2017/FISI\\}
\title{Displaced Vertices from Pseudo-Dirac Dark Matter}
\author[a]{Alessandro Davoli}
\author[a]{Andrea De Simone}
\author[a]{Thomas Jacques}
\author[b]{Ver\'onica Sanz}
\affiliation[a]{SISSA and INFN, via Bonomea 265, 34136 Trieste, Italy}
\affiliation[b]{Department of Physics and Astronomy, University of Sussex, Brighton BN1 9QH, UK\\}

\abstract{ 
Displaced vertices are relatively unusual signatures for dark matter searches at the LHC. We revisit the model of pseudo-Dirac dark matter (pDDM), which can accommodate the correct relic density, evade direct detection constraints, and generically provide observable collider signatures in the form of displaced vertices.
We use this model as a benchmark to illustrate the general techniques involved in the analysis, the complementarity between monojet and displaced vertex searches, and provide a comprehensive study of the current bounds and prospective reach.
}

\emailAdd{alessandro.davoli@sissa.it}
\emailAdd{andrea.desimone@sissa.it}
\emailAdd{thomas.jacques@sissa.it}
\emailAdd{v.sanz@sussex.ac.uk}

\begin{document}
\maketitle
\flushbottom

\section{Introduction}\label{sec:intro}

While we have strong evidence for the existence of dark matter (DM), the search for its particle interactions continues on many fronts. At the forefront of these searches are indirect detection experiments constraining the annihilation of DM, direct detection and solar neutrino experiments constraining the scattering rate, and collider experiments searching for the production of DM. Together these experiments have placed strong constraints on a wide range of DM models, yet a conclusive positive signal remains elusive.

The strength of these constraints is leading to challenges for certain classes of models with relatively strong dark interactions. It can be difficult to find regions of parameter space that lead to the correct relic density while avoiding existing constraints, see for example Refs.~\cite{Duerr:2016tmh,Alves:2015pea} covering a wide range of constraints in the context of simplified models. In particular, constraints on the spin-independent scattering cross-section from experiments such as LUX \cite{Akerib:2016vxi}, PandaX \cite{Tan:2016zwf} and XENON100 \cite{Aprile:2016swn} are particularly strong and rule out the na\"ive relic density couplings in many models. This can be avoided if the crossing symmetry between the various interactions is broken, reducing the scattering rate while leaving the annihilation rate sufficiently large to avoid overclosing the Universe.

One way to accomplish this is with a model known as pseudo-Dirac DM, described in the EFT limit in Ref.~\cite{DeSimone:2010tf}. This model introduces a pair of dark Majorana fermions with a large Dirac mass, split by a small Majorana mass term, the lighter of which is stable and then represents the DM candidate
(for similar situations, also realized in supersymmetric frameworks, see Refs.~\cite{Abel:2011dc,Giudice:2010wb,Fox:2002bu,TuckerSmith:2001hy,Nelson:2002ca}). The scattering rate is suppressed by spin, avoiding strong constraints on the spin-independent scattering cross-section \cite{Akerib:2016vxi,Tan:2016zwf,Aprile:2016swn}. While the direct annihilation rate is also velocity suppressed, the coannihilation rate is unsuppressed. This leads to a sufficiently large effective annihilation rate necessary to produce the correct relic abundance at the time of thermal freezeout. 
An effective field theory analysis requires that the energy scale of the model be much larger than the typical interaction scale, so that the mediator can be integrated out. At LHC energies, this requires heavy mediators, which often require very large coupling strengths in order to give an observable LHC signature. For this reason, it is often useful to move to simplified models. For some recent reviews, see \cite{Abdallah:2015ter,DeSimone:2016fbz,Kahlhoefer:2017dnp}.

In the present work, we extend the model introduced in Ref. \cite{DeSimone:2010tf} by introducing a $Z'$ gauge boson which couples the dark sector to the Standard Model which, if integrated out, gives rise to the effective operators considered there.

The interaction strength necessary to produce the relic density can lead to observable production rates at current or future runs of the LHC. Further, and crucially, the heavier of the dark particles can be produced with an energy and decay length which can lead to observable displaced vertex signals at the LHC. Displaced vertices and disappearing tracks are a striking signal with no standard model (SM) background, and hence a smoking gun signature of new physics  \cite{Strassler:2006im,Strassler:2006ri,Cheung:2009su,Meade:2009mu,Feng:2010ij,Falkowski:2010cm,Meade:2010ji,Meade:2011du,Aad:2012kw,Aad:2013gva,Helo:2013esa,Jaiswal:2013xra,Aad:2014yea,Buckley:2014ika,Cui:2014twa,CMS:2014wda,Aad:2015rba,Clarke:2015ala,Csaki:2015fba,Csaki:2015uza,Curtin:2015fna,Khachatryan:2015jha,Liu:2015bma,Schwaller:2015gea,Aaboud:2016dgf,Aaboud:2016uth,Allanach:2016pam,Coccaro:2016lnz,Khachatryan:2016sfv,Mahbubani:2017gjh, Buchmueller:2017uqu,  Antusch:2017hhu, Izaguirre:2015zva}.

The remainder of the paper is organized as follows. In Section \ref{sec:model} we will describe the pseudo-Dirac DM model and some of its phenomenology. In Section \ref{sec:params} we will describe existing constraints on the model and our choice of parameters, which we use to estimate prospective LHC constraints and discovery possibilities in Section \ref{sec:results}. We conclude in Section \ref{sec:concls}. 

 \section{Model}\label{sec:model}
  
 The starting point is to consider a generic new four-component Dirac fermion $\Psi$ that is a singlet under the SM gauge group. We consider the most general Lagrangian  for $\Psi$ with both Dirac ($M_D$) and Majorana ($m_{L,R}$) masses 
\cite{DeSimone:2010tf}:
\begin{equation}
\mathcal{L}_0=\bar\Psi(i\slashed{\partial}-M_D)\Psi
-\frac{m_L}{2}\left(\overline{\Psi^c} P_L\Psi+\text{h.c.}\right)
-\frac{m_R}{2}\left(\overline{\Psi^c} P_R\Psi+\text{h.c.}\right),
\label{pDDMLfree}
\end{equation}
where $P_{L,R}=(1\mp \gamma^5)/2$.
We focus on the ``pseudo-Dirac'' limit of the mass matrix, 
where $M_D\gg m_L, m_R$.

As an explicit example of the dark sector, we consider it to be completed by a vector mediator $Z'$ of mass $M_{Z'}$. 
The mediator $Z'$ couples $\Psi$ to the Standard Model through renormalizable interactions described by the Lagrangian:
\begin{equation}
\mathcal{L}_{\rm{int}}=\bar\Psi\gamma^\mu(c_L P_L+c_R P_R)\Psi\,
Z_\mu'+
\sum_f\bar f\gamma^\mu(c_L^{(f)} P_L+c_R^{(f)} P_R)f\,Z_\mu'\,,
\label{pDDMLint}
\end{equation}
where $f$ is a SM fermion and $c_{R,L}, c_{R,L}^{(f)}$ are generic operator coefficients which we assume to be real.
We do not commit ourselves to any specific ultraviolet-complete realization of this model of the dark sector, but simply consider it as a simplified phenomenological model. 
Examples of viable ultraviolet completions of this model are the pseudo-Dirac Bino in extended supersymmetry (see discussion in Ref.~\cite{DeSimone:2010tf}), or by considering $Z'$ as a gauge boson of a dark non-abelian gauge group. No dark $U(1)$ completion is possible because the Majorana masses would explicitly break it.

As an explicit example, we could consider the case in which $\Psi$ is embedded within a fermion $\Theta$ which is  a doublet under a (spontaneously broken) SU(2)$^\prime$ hidden gauge symmetry. The Dirac-type mass term for $\Psi$ could then be generated through a Higgs-like mechanism from the vev $v'$ of a heavy scalar field $\Phi'$.

The Majorana-type mass terms, on the other hand, could derive from a Weinberg operator of the form $\frac{1}{\Lambda}\,\bar\Theta\left(i\sigma_2\Phi'\right){\left(i\sigma_2\Phi^\prime\right)}^\dagger\,\Theta^c$, after $\Phi'$ gets a vev. The hierarchy between Dirac and Majorana masses appears to be quite natural, since $M_D\propto v'$ and $m_{L,R}\propto {v'}^2/\Lambda\sim M_D\,v'/\Lambda$, with $\Lambda$ being an effective scale of some underlying high-energy physics.
In the end, the $Z'$ can be viewed as one of the gauge bosons associated with this SU(2)$^\prime$ symmetry.

Such a UV completion turns out to be anomaly-free. Possible anomalies could arise because of the coupling of $Z'$ to SM leptons: in particular, triangle diagrams including U(1)-SU(2)-SU(2)$^\prime$ and U(1)-SU(2)$^\prime$-SU(2)$^\prime$ currents have to be taken into account. The anomalies arising from these diagrams are equal to each other and are proportional to the sum of the hypercharges of the SM  fermions. Therefore, provided that we allow coupling of $Z'$ to all the SM fermions, both  anomalies cancel.
In our analysis, as already mentioned, we focus on the case $f=q$: this means that the couplings to leptons, although effectively present, are vanishingsly small.

The two mass eigenstates, denoted by $\xi_{1,2}$, with masses $m_{1,2}=M_D\mp(m_L+m_R)/2$, will be linear combinations of $\Psi$, $\Psi^c$. It is then possible to construct the Majorana fields (with canonical kinetic term) $\chi_{1,2}$ out of these mass eigenstates: $\chi_1\equiv(\xi_1+\xi_1^c)/2$ and $\chi_2\equiv(\xi_2+\xi_2^c)/2$.
At the zeroth order in $|m_L-m_R|/M_D$, the Majorana eigenstates are given by:
\begin{subequations}
\begin{align}
\chi_1&=\frac{i}{\sqrt 2}\left(\Psi-\Psi^c\right)\label{field_chi1}\\[.1truecm]
\chi_2&=\frac{1}{\sqrt 2}\left(\Psi+\Psi^c\right)\label{field_chi2}\,.
\end{align}
\end{subequations}
The spectrum of this model consists of the lightest state $\chi_1$ with mass $m_1$, identified with a Majorana DM particle, and a slightly heavier companion state $\chi_2$, with mass $m_2$. The model described by the free Lagrangian $\mathcal L_0$ is simply defined by the two mass parameters $m_1$ and $\Delta m\equiv m_2-m_1$ (or, equivalently, $m_1$ and $m_2$). In the pseudo-Dirac limit, the mass splitting satisfies the condition $\Delta m\ll m_1,m_2$.

The free Lagrangian in eq. \eqref{pDDMLfree} then becomes:
\begin{equation}
\mathcal{L}_0=\frac{1}{2}\,\bar\chi_1(i\slashed{\partial}-m_1)\chi_1+\frac{1}{2}\,\bar\chi_2(i\slashed{\partial}-m_2)\chi_2\,.
\label{pDDMLfree2}
\end{equation}
We can then rewrite the interaction Lagrangian in eq.~\eqref{pDDMLint} in terms of $\chi_{1,2}$ as:
\begin{equation}
\mathcal{L}_{\rm{int}}=\mathcal{L}_{\rm{int}}^{(\chi_1\chi_2)}+\mathcal{L}_{\rm{int}}^{(\chi_1\chi_1)}+\mathcal{L}_{\rm{int}}^{(\chi_2\chi_2)}+
\mathcal{L}_{\rm{int}}^{(\bar ff)}\,,
\label{pDDMLint_2}
\end{equation}
where:
\begin{subequations}
\begin{align}
\mathcal{L}_{\rm{int}}^{(\bar ff)}&=\sum_f
\bar f\gamma^\mu\left[\frac{c_L^{(f)}+c_R^{(f)}}{2}-\frac{c_L^{(f)}-c_R^{(f)}}{2}\,\gamma^5\right]f\,Z_\mu'\label{pDDMLintff}\\[.15truecm]\mathcal{L}_{\rm{int}}^{(\chi_1\chi_2)}&=
i\,\frac{c_R+c_L}{2}\,
\bar\chi_1\gamma^\mu\chi_2\,Z_\mu'\label{pDDMLint12}\\[.15truecm]
\mathcal{L}_{\rm{int}}^{(\chi_i\chi_i)}&=
\frac{c_R-c_L}{4}\,
\bar\chi_i\gamma^\mu\gamma^5\chi_i\,Z_\mu'\quad,\quad i=1,2\,.
\label{pDDMLint11}
\end{align}
\end{subequations}

Notice that, remarkably, due to the Majorana nature of the $\chi_i$ fields, the interaction between $\chi_1$ and $\chi_2$ occurs via a pure vector coupling, whereas that between two $\chi_i$'s is a pure axial-vector one. 
These two coupling structures have contrasting phenomenology for scattering and annihilation \cite{Abdallah:2015ter,Kahlhoefer:2015bea,DeSimone:2016fbz,Duerr:2016tmh}. This contrast is one of the core features of the model. Local $\chi_1$ particles scattering with nucleons in the Earth do not have enough energy to upscatter into $\chi_2$, and so scattering proceeds only through $\chi_1 N \rightarrow \chi_1 N$. The axial-vector coupling structure means that this interaction is suppressed by a combination of non-relativistic DM-nucleon scattering operators \cite{DeSimone:2016fbz},
\begin{eqnarray}
\mathcal{O}_4^{\rm NR} &=& \vec{s}_\chi \cdot \vec{s}_N\label{NRopSD}\\
\mathcal{O}_8^{\rm NR} &=& \vec{s}_\chi \cdot \vec{v}_\bot\\
\mathcal{O}_9^{\rm NR} &=& i\vec{s}_\chi \cdot (\vec{s}_N \times \vec{q}),
\end{eqnarray}
where $\vec{s}_{\chi,N}$ is the spin of the DM and nucleon respectively, $\vec{q}$ is the transferred momentum, and $\vec{v}^\bot \equiv \vec{v} - \vec{q} / 2\mu_N$ with $\vec{v}$ the relative velocity and $\mu_N$ the reduced mass of the DM-nucleon system. Each of these are strongly suppressed relative to the spin-independent scattering rate \cite{Fan:2010gt,Fitzpatrick:2012ix,DelNobile:2013sia,Dent:2015zpa}, such that  the model evades strong constraints from direct detection \cite{Akerib:2016vxi,Tan:2016zwf,Aprile:2016swn}.

The axial-vector interaction usually leads to a suppressed annihilation rate, such that very large couplings would be required to produce the correct relic abundance. The presence of an unsuppressed vector coannihilation term alleviates this problem, as discussed in Section~\ref{sec:relics}.

\subsection{Decay length\label{sec:decay_length}}
The expressions for the interaction Lagrangian in eqs. \eqref{pDDMLintff} and \eqref{pDDMLint12} are responsible for the decay $\chi_2\to\chi_1f\bar f$; the decay width for this process at leading order in the small parameters $\Delta m/m_1$ and $m_f/m_1$ is given by
\begin{equation}
\Gamma_{\chi_2\to\chi_1\bar ff}\simeq\sum_f\frac{N_c^{(f)}}{480\pi^3}\,{\left(c_L+c_R\right)}^2\left({c_L^{(f)}}^2+{c_R^{(f)}}^2\right)\frac{\Delta m^5}{M_{Z'}^4}\,,
\label{decay_width_estimate}
\end{equation}
where $N_c^{(f)}$ is then number of colours of the fermion $f$.
A more general expression is reported in eq.~\eqref{decay_width_chi_2}.
In the present work, we focus our attention on quarks, but the formula above can be applied to a generic Standard Model fermion.

The previous equation \eqref{decay_width_estimate} also allows the determination of the decay length of $\chi_2$; in particular, if it decays at rest, the mean decay length is simply $L_0=1/\Gamma_{\chi_2\to\chi_1\bar ff}$.
The decay length at rest corresponding to eq.~\eqref{decay_width_estimate} is
\begin{equation}
L_0\simeq \SI{2.94}{\meter} \,{\left[\displaystyle\sum_f N_c^{(f)}{\left(c_L+c_R\right)}^2\left({c_L^{(f)}}^2+{c_R^{(f)}}^2\right)\right]}^{-1}{\left(\frac{M_{Z'}}{\SI{1}{\tera\electronvolt}}\right)}^4 {\left(\frac{\SI{1}{\giga\electronvolt}}{\Delta m}\right)}^5\,.
\label{estimate_L_0}
\end{equation}
The corrections proportional to $\Delta m/m_1$ and $m_f/m_1$ can be of the order of 30\%, but eq.~\eqref{estimate_L_0} correctly reproduces the order of magnitude of such a decay length. 
In particular, it shows that for a mass splitting of $\mathcal O(\SI{}{\giga\electronvolt})$, and mediator mass of $\mathcal O(\SI{}{\tera\electronvolt})$, the decay length can be of the order of the radius of the ATLAS and CMS detectors, allowing the observation of a displaced vertex signal. 
Since in the following we will  mainly be interested in studying this decay in a collider, the formula above must be corrected to include the boost factor for $\chi_2$; this translates into a mean decay length in the laboratory frame given by:
\begin{eqnarray}
L_0^\mathrm{lab}=\beta\gamma L_0\,,
\end{eqnarray}
where $\beta\gamma\equiv p_2/m_2$ is the boost factor for $\chi_2$. The decay length $ L_{\rm lab}$ of a particle in the detector with a given momentum then follows the probability distribution 
\begin{equation}
P(L^{\rm lab}) = \frac{1}{L_{0}^{\rm lab}} \,e^{-L^{\rm lab}/L_{0}^{\rm lab}}.
\end{equation}
We can define a decay length in the transverse direction of the detector as $L_{T,0}^\mathrm{lab} \equiv L_0\,p_2^T/ m_2$ where $p_2^T$ is the $\chi_2$ momentum in the transverse direction. %
Following Ref.~\cite{ElHedri:2017nny}, the final probability of the transverse decay length being greater than some length $L$, after integration over the probability  distributions of the kinematic variables, can be closely approximated by simulating and averaging over a large number of events $N$, 
\begin{equation}
P(L_T^{\rm lab} > L) = \frac{1}{N} \sum_{i = 1}^N\exp\left(-\frac{L}{L_{T,0}^{\rm lab}(p_2^T = p_{2,i}^T)}\right)\,.
\end{equation}
%

\subsection{Relic abundance\label{sec:relics}}
The model we are considering is characterized by a mass splitting which in general satisfies the condition $\Delta m\ll m_{1,2}$; this means that the two states are quasi-degenerate, and coannihilations are therefore important in the determination of the correct relic abundance. As we will see, coannihilations are especially relevant in this model given that $\chi_i\chi_i$ annihilations are generally suppressed relative to coannihilations, with some dependence on the choice of couplings. In particular, the effective cross-section is given by \cite{Griest:1990kh}:
\begin{equation}
{\langle\sigma v\rangle}_\mathrm{eff}=\frac{1}{{\left(1+\alpha\right)}^2}\left({\langle\sigma v\rangle}_{11}+2\alpha{\langle\sigma v\rangle}_{12}+\alpha^2{\langle\sigma v\rangle}_{22}\right)\,,
\label{sigma_eff}
\end{equation}
where we have defined $\alpha\equiv{(1+\Delta m/m_1)}^{3/2}e^{-x\Delta m/m_1}$, $x\equiv m_1/T$ and ${\langle\sigma v\rangle}_{ij}\equiv{\langle\sigma v\rangle}_{\chi_i\chi_j\to f\bar f}$.

For the interactions in eqs. \eqref{pDDMLintff}-\eqref{pDDMLint11}, the effective thermal cross-section is, with the same approximations made to obtain eq. 
\eqref{decay_width_estimate}: 
%
\begin{equation}
\langle\sigma v\rangle_{\mathrm{eff}}\simeq\sum_f\frac{N_c^{(f)}}{16\pi}{\left(c_L+c_R\right)}^2\left({c_L^{(f)}}^2+{c_R^{(f)}}^2\right)\frac{m_1^2}{M_{Z'}^4}\,.
\label{approx_sigma_eff}
\end{equation}
A numerical estimate gives:
\begin{equation}
\frac{\langle\sigma v\rangle_{\mathrm{eff}}}{\langle\sigma v\rangle_{\mathrm{WIMP}}}\simeq0.08\sum_fN_c^{(f)}{\left(c_L+c_R\right)}^2\left({c_L^{(f)}}^2+{c_R^{(f)}}^2\right){\left(\frac{m_1}{\SI{100}{\giga\electronvolt}}\right)}^2{\left(\frac{\SI{1}{\tera\electronvolt}}{M_{Z'}}\right)}^4\,,
\label{estimate_sigma_eff}
\end{equation}
where $\langle\sigma v\rangle_{\mathrm{WIMP}}\equiv\SI{3e-26}{\cubic\centi\meter\per\second}$ is the typical WIMP annihilation cross-section.

Even in this case, this is just an estimate: more complete expressions, including corrections proportional to quark masses, are reported in  eqs. \eqref{sigma_12} and \eqref{sigma_ii}.

It is important to notice that the $\chi_i\chi_i$ self-annihilations are velocity suppressed, whereas the coannihilation $\chi_1\chi_2$ is not (cf. eqs.  \eqref{sigma_12} and \eqref{sigma_ii}). Nonetheless, due to the different dependence on couplings, both  terms should be kept in the determination of the effective thermal cross-section.

The relic abundance is then related to the effective cross-section as
\begin{equation}
\Omega h^2=\frac{\SI{8.7e-11}{\per\giga\electronvolt\squared}}{\sqrt{g_*}\displaystyle\int_{x_F}^\infty dx\,\frac{{\langle\sigma v\rangle}_\mathrm{eff}}{x^2}}\,,
\label{relic_abundance}
\end{equation}
where $g_*$ is the number of relativistic degrees of freedom at the freeze-out temperature $T_F$, determined by the implicit equation
\begin{equation}
x_F=25+\log\left[\frac{d_F}{\sqrt{g_*x_F}}\,m_1{\langle\sigma v\rangle}_\mathrm{eff}\,\SI{6.4e6}{\giga\electronvolt}\right]\,,
\end{equation}
with $d_F$ being the number of degrees of freedom of the $\chi_i$'s, $d_F=2$ in the present model. In the following, we take $g_*=96$.

\subsection{Link between decay length and relic abundance}

It is remarkable to notice 
that the approximate expression in eq.~\eqref{decay_width_estimate} and the $s$-wave contribution in eq.~\eqref{estimate_sigma_eff}
contain the same combination of couplings. This is a consequence of the fact that the same matrix element controls the decay of $\chi_2 \to \chi_1 \bar f f$ and the co-annihilation $\chi_1\chi_2 \to \bar f f$. 

In the limit of massless SM fermions $m_f=0$, the self-annihilations are velocity-suppressed and therefore the relic abundance is  dominated by the co-annihilations.
This way, a very intriguing link can be traced between a cosmological property (relic density) and a collider observable (decay length), as already noticed in Ref.~\cite{DeSimone:2010tf}.

The combination of couplings entering the decay length can then be traded for the (known) relic abundance, thus establishing a very direct correlation between the decay length $L_0$, the DM mass $m_1$ and the mass splitting $\Delta m$. 
By combining eqs. \eqref{estimate_L_0}, \eqref{estimate_sigma_eff} and \eqref{relic_abundance}, we can write the relic abundance as a function of $L_0$ as:
\begin{equation}
\frac{\Omega h^2}{0.1194}\simeq1.26\;\frac{x_F}{\sqrt{g_*}}\,\frac{1}{1+\dfrac{1}{2x_F}{\left(\dfrac{1-k}{1+k}\right)}^2}{\left(\frac{L_0}{\SI{1}{\meter}}\right)}{\left(\frac{\SI{100}{\giga\electronvolt}}{m_1}\right)}^2{\left(\frac{\Delta m}{\SI{1}{\giga\electronvolt}}\right)}^5\,,
\label{estimate_Omega}
\end{equation}
with $k\equiv c_R/c_L$.

From the equation above, we can estimate the value for $L_0$ for given $(m_1,\Delta m,k)$ by imposing the measured value for $\Omega h^2$. In addition, we see that for given $L_0$, eq.~\eqref{estimate_Omega} does not depend on $M_{Z'}$, and since  $x_F\sim\mathcal O(20)$, then if $k\gtrsim0$, it depends only very mildly on $k$.

If one is able to infer  $L_0$ (from the displaced vertex) and $\Delta m$ (from the edge of di-jet or di-lepton distribution) by collider measurements, then it would be possible to make a prediction for the DM mass $m_1$.

 \section{Constraints and Choice of Parameters}\label{sec:params}
The model has a parameter space spanned by seven parameters:
\begin{equation}
\{m_1,m_2,M_{Z'},c_L, c_R, c_L^{(f)}, c_R^{(f)}\}.
\end{equation}
In order to avoid a full scan over the entire seven-dimensional parameter space, we can motivate benchmark points and apply a number of constraints before performing the main analysis. We will leave $\{m_1,m_2\}$ free, which we will usually parameterise as $\{m_1, \Delta m\}$. 
Our signals of interest are not sensitive to the chirality of the quarks, and so without loss of generality, we can set $c_R^{(f)}=-c_L^{(f)}$. This leads to a pure axial-vector coupling between the $Z'$ and SM quarks. 
We have checked that perturbative unitarity is not violated for the values of masses and couplings considered in our analysis \cite{Kahlhoefer:2015bea}.

In this situation, the non-relativistic DM-nucleon scattering operator is given by eq.~\eqref{NRopSD}, which leads to a pure `spin-dependent' scattering cross-section, such that constraints from direct detection constraints on $\sigma_{\rm SD}$ can be applied directly using \cite{Boveia:2016mrp}:
\begin{equation}
\sigma_{\rm SD} \simeq\SI{2.4e-42}{\square\centi\meter} \cdot (c_R - c_L)^2 \left(c_L^{(f)}\right)^2 \left(\frac{\SI{1}{\tera\electronvolt}}{M_{Z'}}\right)^4 \left(\frac{\mu_{n\chi}}{\SI{1}{\giga\electronvolt}}\right)^2,
\end{equation}
in the case where $c_L^{(f)} = -c_R^{(f)}$, and $\mu_{n\chi}$ is the DM-nucleon reduced mass. We find that current direct detection limits such as from LUX \cite{Akerib:2017kat} are substantially weaker than other constraints, and play no further role in the determination of the couplings below.

The relative contributions of the axial-vector $\chi_i \chi_i$ and vector $\chi_1 \chi_2$ coupling is controlled by the ratio $k=c_R/c_L$. The axial-vector term is proportional to $|c_R - c_L|$, and so in the limit $k \rightarrow 1$, the $\chi_i \chi_i$ term disappears. Conversely, the limit $k \rightarrow -1$, the vector term disappears and the decay length increases as seen in eq.~\eqref{estimate_L_0}. The interplay between these two contributions is important for the potential observability of displaced vertices, and so we choose two benchmarks for $k$ showing different regions of phenomenology, specifically $k=-0.8$ and $k=0$. 

Note that a degeneracy arises because in all relevant observables, $c_{L,R}$ appear together as either $|c_L + c_R|^2$  or $|c_L - c_R|^2$. Therefore the ($c_L$,$c_R$) plane is divided into 4 equivalent wedges separated by the lines defining $k =-1$, $k=1$. Any point in one of the 4 wedges can be mapped onto a point in any of the other 4 wedges with no change in the phenomenology. The consequence of this is a degeneracy such that choosing $k=c_R/c_L = -0.8\,(0)$ is equivalent to choosing $c_L/c_R = -0.8\,(0)$. Similarly the transformation $(c_R, c_L) \rightarrow (-c_R, -c_L)$ has no effect.

In the following  subsection, we will discuss dijet constraints which strongly restrict the $Z'$ couplings to quarks $c_L^{(f)}, c_R^{(f)}$. 
Next we will require that the model reproduces the correct relic density, breaking the degeneracy by restricting us to a contour of $c_L$ and $c_R$, and leaving us with a full set of benchmark parameter choices.
Finally we will impose the requirement that the width of the $Z'$ remains modest, which restricts the parameters to remain within a contour of $c_L$ and $c_R$.

\subsection{Dijets}
Dijet searches put upper bounds on the couplings between the Standard Model and the dark mediator.

In the following, we take the results of Ref. \cite{ATLAS:2016lvi}: in particular, in their Fig.~4, limits on the coupling between $Z'$ and SM quarks in an axial-vector simplified model are shown. These constraints derive from a limit on the mediator production rate scaled by the branching ratio into quarks, and is hence sensitive to the ratio between the DM coupling and the quark coupling.  Ref. \cite{ATLAS:2016lvi} assumes a negligible coupling to DM, which provides the strongest possible limits. Including a fixed coupling to DM would decrease the branching fraction to quarks and hence weaken the constraints. We choose not to apply this rescaling, which would allow larger values of $c_L^{(f)},\, c_R^{(f)}$, in order to be conservative and to be consistent with possible future constraints.

The constrained parameter $g_q$ of Ref.~\cite{ATLAS:2016lvi}  is equivalent to our parameter $c_L^{(f)}$ given that we are considering the case $c_L^{(f)} = -c_R^{(f)}$.
We consider three benchmark values for $M_{Z'}$; for each, we select a benchmark value for $c_L^{(f)}$, chosen to be large while still compatible with the bounds of Ref.~\cite{ATLAS:2016lvi}. These choices are shown in Table~\ref{tab:value_params}.
We choose these couplings to be universal, i.e. to be the same for all quarks and to be independent of the value of $m_1$ and $\Delta m$.


\subsection{Relic density}
\label{subsec:relic_density}
\begin{figure}[t]
\centering
\includegraphics[width = 0.45\linewidth]{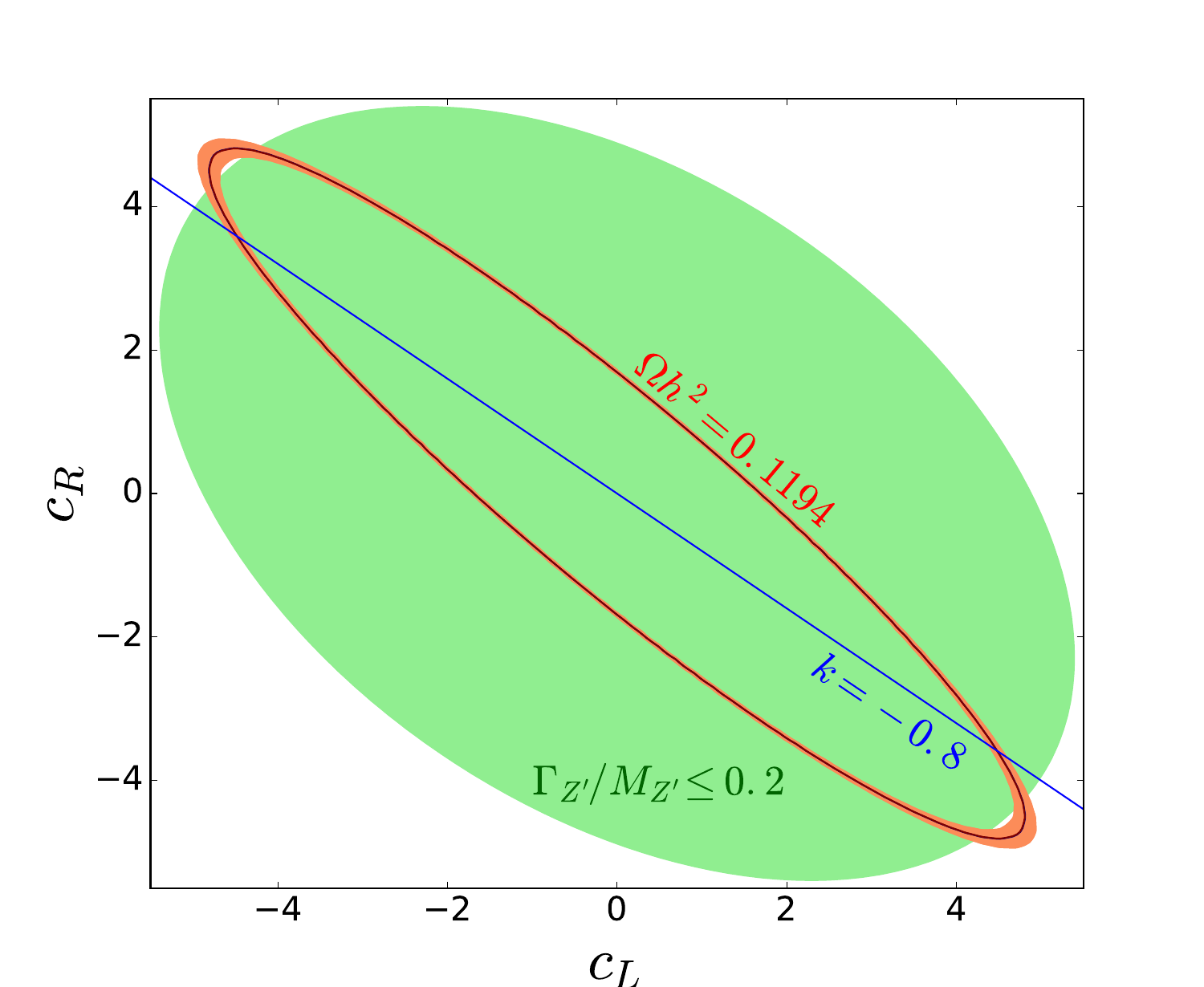}
\includegraphics[width = 0.45\linewidth]{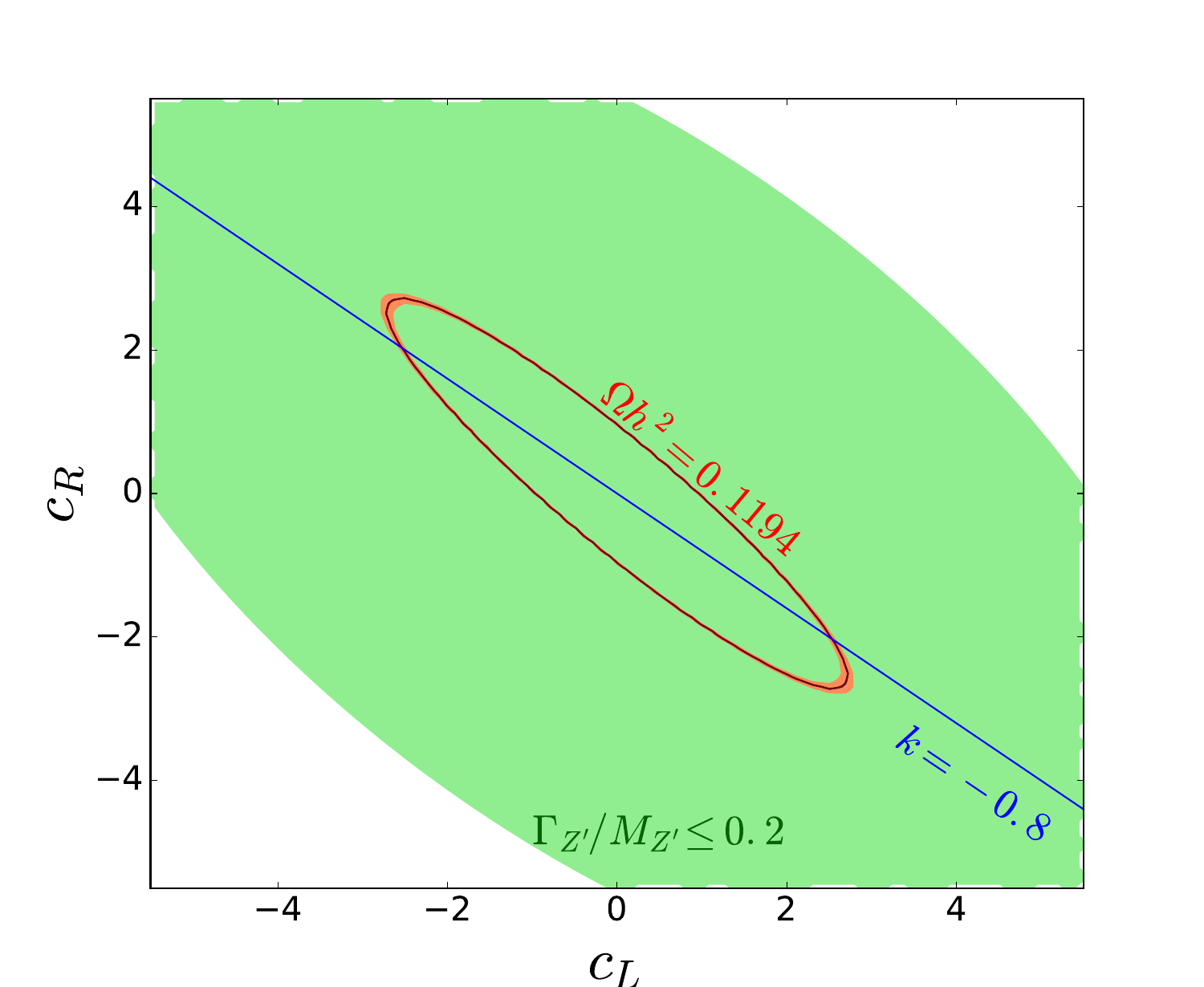}
\includegraphics[width = 0.45\linewidth]{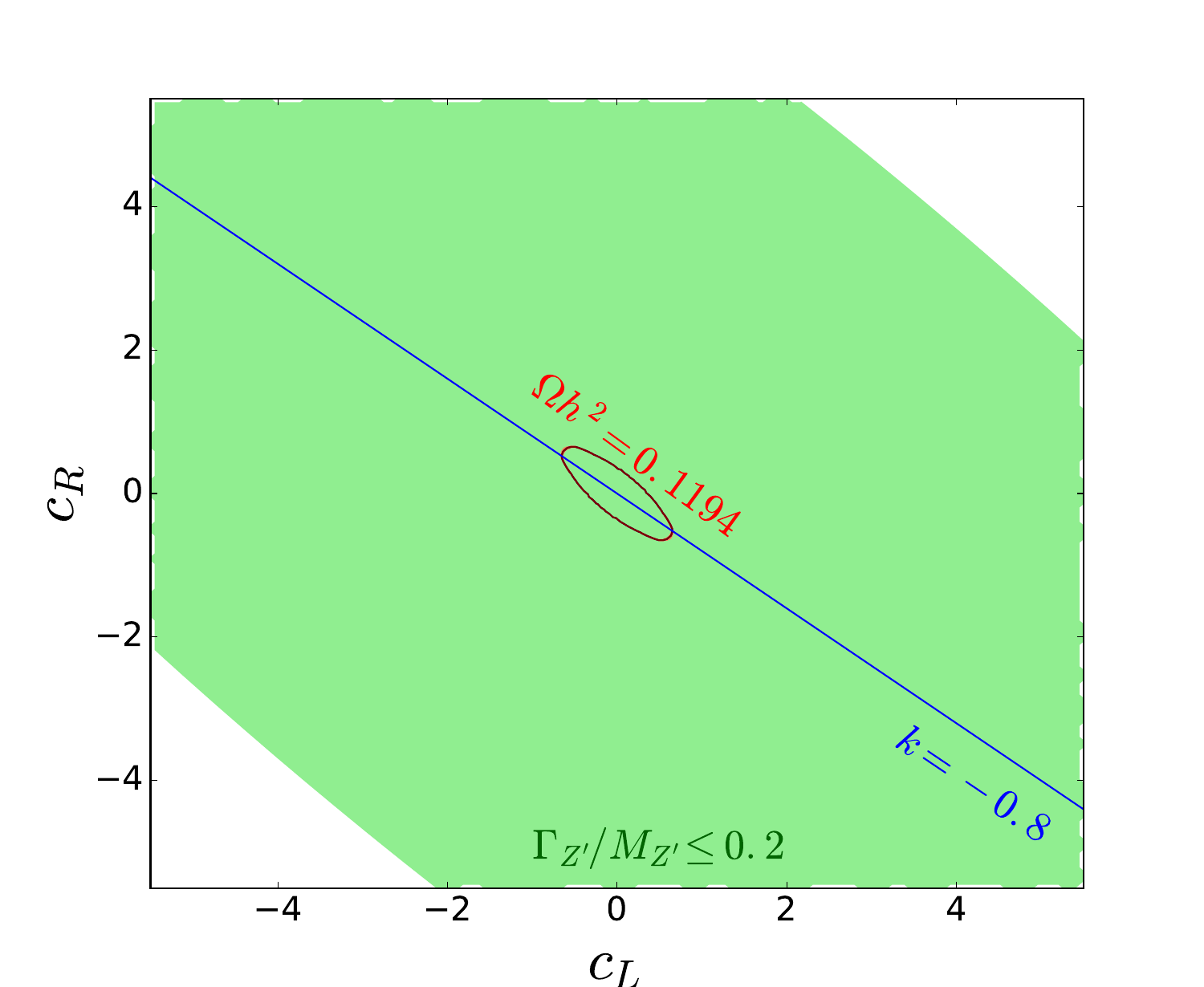}
\caption{\label{fig:plane_restriction}Interplay between the region $\Gamma_{Z'}/M_{Z'}\leq0.2$ (green region) and observed dark matter abundance \cite{Ade:2015xua} $\Omega h^2=0.1194$ (black line with orange contours), for $M_{Z'}=\SI{1.5}{\tera\electronvolt}$, $\Delta m=\SI{5}{\giga\electronvolt}$, $k=-0.8$ (blue line) and $c_L^{(f)}=0.07$, for different values of $m_1$: $m_1=\SI{525}{\giga\electronvolt}$ (left-top panel), $m_1=\SI{610}{\giga\electronvolt}$ (right-top panel), $m_1=\SI{700}{\giga\electronvolt}$ (bottom panel). The orange contours correspond to $3\sigma$ deviations from the best value.}
\end{figure}

For given values of $m_1$, $\Delta m$ and $M_{Z'}$,  we can determine the contour in the $(c_L,c_R)$ plane which corresponds to the observed DM relic abundance using eq. \eqref{relic_abundance}.  
%
For the observed value, we take $\Omega h^2=0.1194$ \cite{Ade:2015xua}. This contour is shown as a black (with orange contour) line in Fig.~\ref{fig:plane_restriction} for different values of $m_1$.

The benchmark choices made earlier for $k=c_R/c_L$ identify a straight line in this plane, shown as a blue line in Fig.~\ref{fig:plane_restriction}, which intercepts the relic abundance contour at two points. 
Recall from the start of this section that the phenomenology of the model is equivalent under the transformation $(c_R \rightarrow -c_R, c_L \rightarrow -c_L)$ (and also under the transformation $k \rightarrow 1/k$). For each value of $m_1$, $\Delta m$ and $M_{Z'}$, and with $c_L^{(f)}$, $c_R^{(f)}$ chosen as described in the previous subsection, the intercept defines the value of $c_L$ and $c_R$, where we make the arbitrary choice $c_R \geq 0$, $c_L < 0$. 

%

\subsection{$Z'$ width}
A final restriction on $c_L$ and $c_R$ comes from a kinematic argument, by imposing the condition $\Gamma_{Z'}\ll M_{Z'}$ in order for our treatment of $Z'$ as a physical particle appearing in the $s$-channel to be consistent.


In Appendix \ref{appendix_1}, we provide explicit expressions for the partial widths of the $Z'$ boson. The ratio of the width approximately goes like $\Gamma_{Z'}/M_{Z'} \appropto (c_L^2 + c_R^2)$, and requiring that this ratio remains below some maximum value defines an oval allowed region in the $c_L, c_R$ plane for a given choice of $m_1, \Delta m$ (now that $c_L$ is fixed by dijet constraints). We set this ratio as $\Gamma_{Z'}/M_{Z'} < 0.2$, above which the Breit-Wigner approximation to the width begins to break down \cite{An:2012va,Jacques:2015zha}. This allowed region is shown in green in Fig.~\ref{fig:plane_restriction}.

For a given choice of $M_{Z'}$, $\Delta m$ and $c_L^{(f)}$, this restricts us to a fixed range of values of $m_1$; as can be seen in Fig.~\ref{fig:plane_restriction}, below a minimum value for $m_1$, the intercept between the relic density contour and $k$ benchmark is outside the green region.
The mass ranges we consider are shown in Tables~\ref{tab:value_params} and ~\ref{tab:value_params_2}, for different values of $k$. 
We choose the same range for $\Delta m$ for all values of $M_{Z'}$, i.e. $\SI{1.5}{\giga\electronvolt}\leq\Delta m\leq\SI{8.0}{\giga\electronvolt}$.

In this way, we have uniquely determined the values of all the couplings, allowing us to find a set which is compatible with both the $Z'$ width and current cosmological observations.

\begin{table}[t]
\begin{center}
\begin{tabular}{c||c|c|c}
$$&$M_{Z'}=\SI{1.5}{\tera\electronvolt}$&$M_{Z'}=\SI{2.5}{\tera\electronvolt}$&$M_{Z'}=\SI{3.5}{\tera\electronvolt}$\\\hline\hline
$c_L^{(f)}$&$0.07$&$0.13$&$0.25$\\\hline
$m_{1,\mathrm{min}}\,(\SI{}{\giga\electronvolt})$&$525$&$850$&$1100$\\\hline
$m_{1,\mathrm{max}}\,(\SI{}{\giga\electronvolt})$&$700$&$1200$&$1600$
\end{tabular}
\end{center}
\caption{\label{tab:value_params}Allowed range of $m_1$ and choice for $c_L^{(f)}$, for different values of $M_{Z'}$ and $k=-0.8$ (equivalent to $k^{-1}= -0.8$).}
\begin{center}
\begin{tabular}{c||c|c|c}
$$&$M_{Z'}=\SI{1.5}{\tera\electronvolt}$&$M_{Z'}=\SI{2.5}{\tera\electronvolt}$&$M_{Z'}=\SI{3.5}{\tera\electronvolt}$\\\hline\hline
$c_L^{(f)}$&$0.07$&$0.13$&$0.25$\\\hline
$m_{1,\mathrm{min}}\,(\SI{}{\giga\electronvolt})$&$375$&$550$&$650$\\\hline
$m_{1,\mathrm{max}}\,(\SI{}{\giga\electronvolt})$&$700$&$1200$&$1600$
\end{tabular}
\end{center}
\caption{\label{tab:value_params_2}Allowed range of $m_1$ and choice for $c_L^{(f)}$, for different values of $M_{Z'}$ and $k=0$ (equivalent to $k^{-1}= 0$).}
\end{table}


 \section{Analysis and Results}\label{sec:results}

So far we have discussed the region of parameter space to be used for the LHC analyses, by imposing a series of constraints. In this section we describe the complementarity between monojet searches and displaced vertex signatures. Searches for pseudo-Dirac DM can be initiated by triggering on events with a single high-$p_T$ jet, with displaced signatures becoming apparent during the offline reconstruction.

We start the section by describing the current 13 TeV monojet analysis, obtaining the current exclusions and estimating the future reach, before moving on to the displaced vertex signatures. These two types of searches are complementary, sensitive to different SM backgrounds and with potentially different scalings at high-luminosity. For the pseudo-Dirac DM model, monojet could provide the first hint of new physics, while the displaced vertex analysis could be used to characterize such an excess as originating from a DM scenario.

\subsection{Monojet analysis}
\label{sec:monojet}
Searches for new physics in events with an energetic jet and a large amount of transverse energy have been performed by ATLAS and CMS. In this section we use the results from the 13 TeV data by ATLAS ~\cite{Aaboud:2016tnv} with 3.2 fb$^{-1}$ to exclude part of the parameter space of the model as well as to obtain projections for higher-luminosity runs. 
 
The production of the stable $\chi_1$ particle can be explored using monojet events where the jet is radiated from the initial state.  Moreover, in the region relevant for dark matter, the associated production of $\chi_1$ with $\chi_2$ and subsequent decay of $\chi_2$ into jets would also lead to monojet signatures. This is a situation complementary to the one which will be described in the next section, where the decay of $\chi_2$ into jets with a displaced signature will be exploited. As discussed there, there is a region of the parameter space where the $\chi_2$ decay appears as prompt. To capture these two topologies, we propose a projected analysis of monojet events at LHC13 with high-luminosity, along the lines discussed in  Ref.~\cite{Barducci:2015ffa}.

We have simulated the processes
\bea
p p &\to & \chi_{1,2}\, \chi_{1,2}\, j , \, \textrm{ with } \chi_2 \to \chi_1 \, j j
\eea
in the range of masses and couplings defined in Table~\ref{tab:value_params} and ~\ref{tab:value_params_2}. We have then applied the selection cuts described in the ATLAS search described in \cite{Aaboud:2016tnv} to determine the current constraints on the parameter space. The experimental search is separated in seven signal regions IM1-IM7, with cuts on missing energy ranging from 250 GeV to 700 GeV. To obtain current exclusions we used the bound on the value of the cross-section at 95\% CL provided, $\langle \sigma \rangle^{95}_{obs}$, which ranged from 553 fb to 19 fb in the IM1 and IM7 regions.  

The constraints do depend on the choices of the parameter $k$, and for $k=-0.8$, only the point of $m_1$= 525 GeV for $M_{Z'}$ =1.5 TeV is ruled out, whereas for $k=0$ a larger region of the parameter space is excluded by this dataset. Indeed, in this case for $M_{Z'}$=(1.5, 2.5, 3.5)  TeV, the region below (550, 800, 850) GeV does not survive the monojet constraints.  
It may appear counterintuitive that for heavier $M_{Z'}$ the monojet excludes larger values of the DM mass; however, the selection procedure described in Section \ref{sec:params} calls for larger couplings as $M_{Z'}$ increases. The net effect is that the signal strength remains approximately constant.

 


The next step is to obtain projections for higher luminosities. To produce the projections, we have to estimate the uncertainties on the SM backgrounds at a given luminosity. Those backgrounds are mainly $ Z j\to \nu\bar\nu j $ and $W j\to l\nu_l j$. In Ref.~\cite{Barducci:2015ffa} a simulation of the main backgrounds was performed and used to project exclusions,  but a more accurate estimate can be obtained by examining the details in the ATLAS analysis. There systematic uncertainties were given, ranging from 2\% in IM1 to 4\% in IM7, as well as the number of expected events at 3.2 fb$^{-1}$ (which can be scaled up to other luminosities). 

To give an example, one could use these numbers to estimate the SM background events at 100 fb$^{-1}$ as 5220 $\pm$ 210 in the region IM7, where we have assumed systematic uncertainties dominate and remain of the same order as in the current analysis. One could then assume the number of observed events to be compatible with the background expectation, and use this to set a 95\% CL limit on the new physics cross-section $\langle \sigma \rangle^{95} \simeq$ 4 fb.

It is possible that at high-luminosity a better control on systematics is achieved or, on the contrary, the high-luminosity environment could lead to a degradation of the understanding of the SM backgrounds. For illustration purposes, we adopt a benchmark choice of $\langle \sigma \rangle^{95}$= 5 fb, which corresponds to a total uncertainty on the SM backgrounds of 10\%. The results are shown in Fig.~\ref{fig:dv_limits} for the cases $k=0$ and $k=-0.8$, together with the displaced vertices contraints discussed in the next subsection. The exclusion limit is roughly independent of $\Delta m$ as the monojet cuts select mostly events with a jet coming from initial state radiation.

\subsection{Displaced Vertices \label{sec:dv}}

\begin{figure}[t]
\centering
\includegraphics[width=0.4\linewidth]{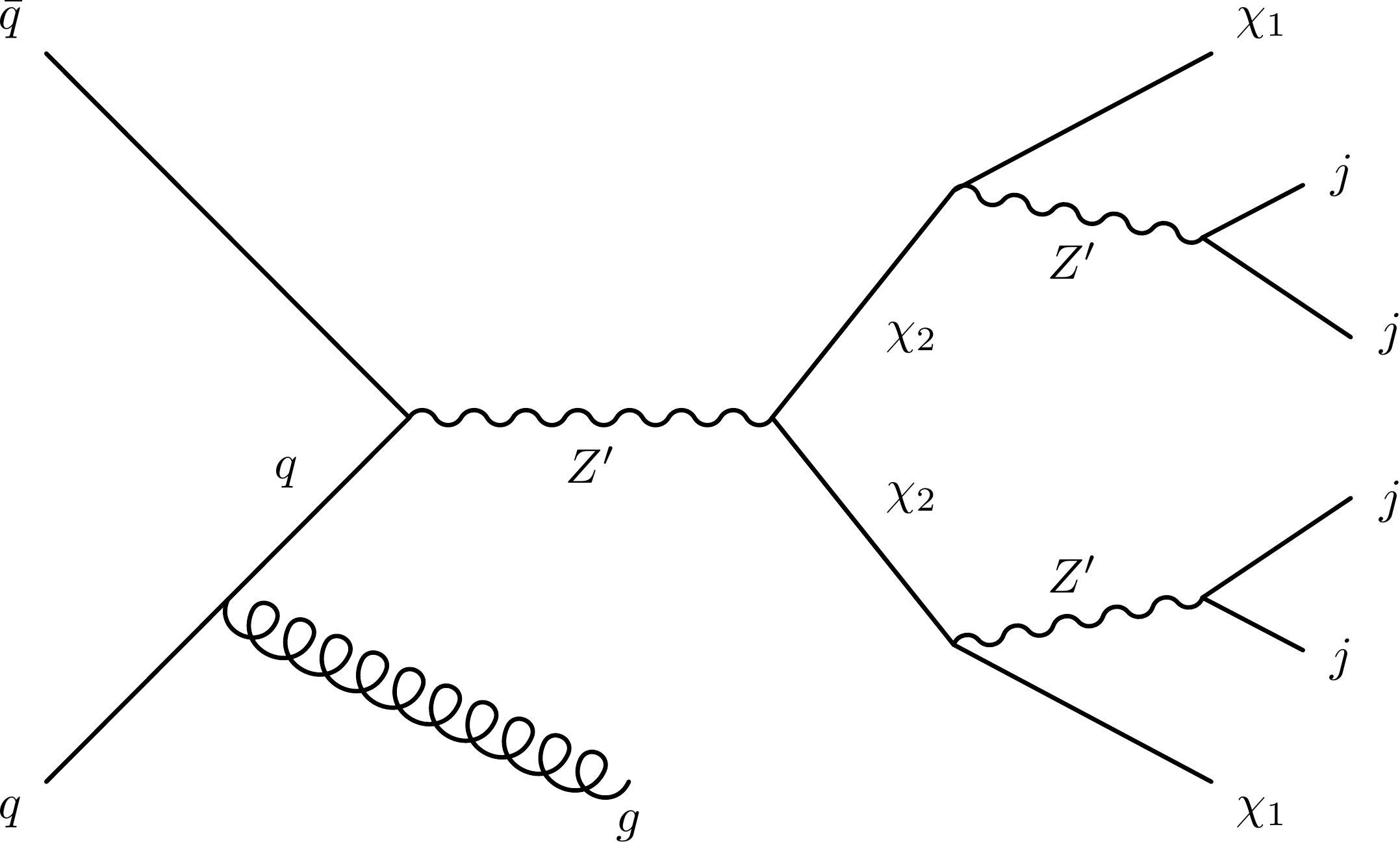}
\caption{\label{fig:diagram}Illustrative Feynman diagram for the displaced vertex process of interest, including the decay of $\chi_2$. Initial state can also be quark-gluon with an ISR quark jet.}
\end{figure}

Displaced vertices are a strong signal of beyond-Standard-Model physics, with a low expected background arising solely from vertex misidentification. The pDDM model predicts a displaced vertex signal at the LHC from $\chi_2$ decay within the detector volume into a $\chi_1 j j$ final state. The strongest signals are expected from the process $p p \rightarrow \chi_2 \chi_2 j \rightarrow \chi_1 \chi_1 5j$, shown in Fig.~\ref{fig:diagram}: the production of two $\chi_2$ particles can lead to two displaced vertices, which has an extremely small expected background, and the emission of initial state radiation (ISR) pushes the $\chi_2$ particles out of a back-to-back configuration, increasing the missing energy signal and allowing us to trigger on events with a high-$p_T$ jet plus missing energy. 
Since we are interested in the region of parameter space with $\Delta m < 10$ GeV, the jets from decay of $\chi_2$ associated with the displaced vertices have $p_T \sim \mathcal{O}(\SI{1}{\giga\electronvolt})$ and are therefore too soft to trigger on, but can be used for the offline analysis and identification of the displaced vertices \cite{Aad:2015rba}.
We simulate at the parton level using the method outlined in Appendix~\ref{appendix_2}.

\begin{figure}[t]
\centering
\includegraphics[scale=0.65]{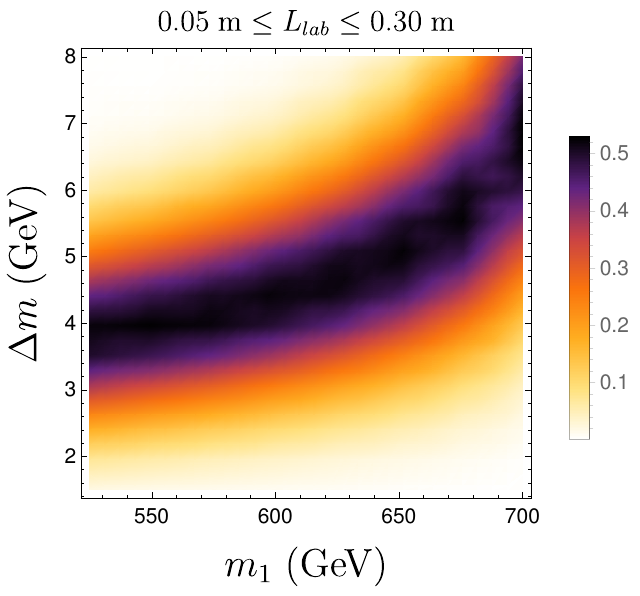}
\includegraphics[scale=0.65]{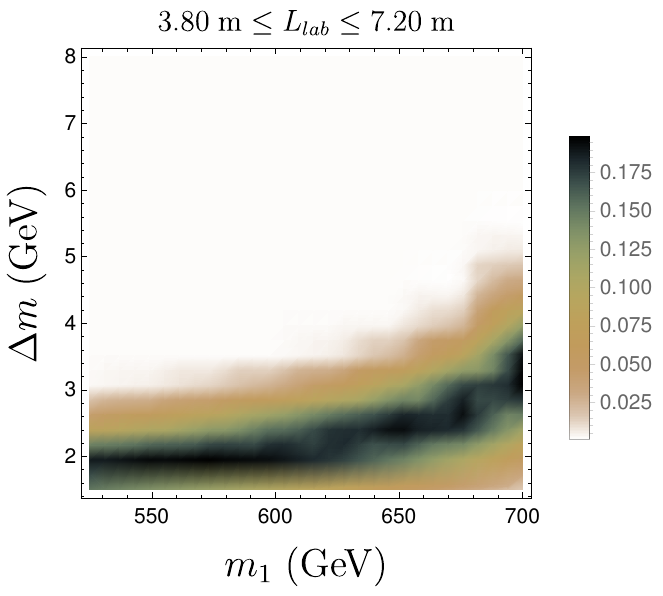}
\caption{\label{fig:probability_decay}Probability that $\chi_2$ decays in either the ATLAS innerd detector (left panel) or muon solenoid (right panel), for $M_{Z'}=\SI{1.5}{\tera\electronvolt}$ and $k=-0.8$.}
\end{figure}

Using the method described in Section \ref{sec:decay_length}, we can compute the probability that $\chi_2$, produced in a $pp$ collision, decays with a decay length within the range of the ATLAS inner detector or muon solenoid.  
We consider the inner detector with radius $r$ defined by $\SI{0.05}{\meter}< r <\SI{0.3}{\meter}$ and the muon solenoid between $\SI{3.8}{\meter}< r <\SI{7.2}{\meter}$, based on the range of displaced vertex identification efficiency from Ref.~\cite{Aad:2015uaa}.
Since the couplings are uniquely fixed as described in Section \ref{sec:params}, such a decay length is a function of $\{m_1,\Delta m,M_{Z'}\}$ only. 
In Fig.~\ref{fig:probability_decay}, we show the result for $M_{Z'}=\SI{1.5}{\tera\electronvolt}$.

We apply the constraints on this process from Ref.~\cite{Aad:2015uaa} by the ATLAS collaboration, which places limits on the number of events with two displaced vertices at center of mass energy $\SI{8}{\tera\electronvolt}$ using a range of selection criteria. Given that our process has large jet $p_T$ and large missing energy, the best limits on our process come from the jets + missing energy trigger, which allows for topologies with displaced vertices in both the muon solenoid and inner detector. The background found by the ATLAS collaboration turns out to be $<10^{-4}$ at  $\sqrt{s}=\SI{8}{\tera\electronvolt}$ with $\SI{20.3}{\per\femto\barn}$ of data, with cuts of leading jet $p_T > \SI{120}{\giga\electronvolt}$, $\mathrm{MET}>\SI{200}{\giga\electronvolt}$. In order to ensure that the expected background remains approximately zero at $\SI{13}{\tera\electronvolt}$, we scale these  cuts on our signal process to $p_T >\SI{200}{\giga\electronvolt}$ and $\mathrm{MET}>\SI{300}{\giga\electronvolt}$. The strong jet $p_T$ and missing energy cuts mean that pseudorapidity $\eta$ is small and no events are found in the barrel endcap.

The jets + MET trigger requires at least 7 tracks per vertex. Whilst a full detector simulation and evaluation of the efficieny is beyond the scope of this paper, we have performed a Delphes-level \cite{deFavereau:2013fsa} analysis of the process for several benchmark points in parameter space and found that approximately 25\% to 50\% of vertices passed this track requirement. With this in mind, the 20\% vertex identification efficiency we use can be considered an optimistic scenario for near-future displaced-vertex experimental analyses, and further emphasises the need for an increased focus on this signal by ATLAS and CMS.

\begin{figure}[thb]
\centering
\includegraphics[width = 0.425\linewidth]{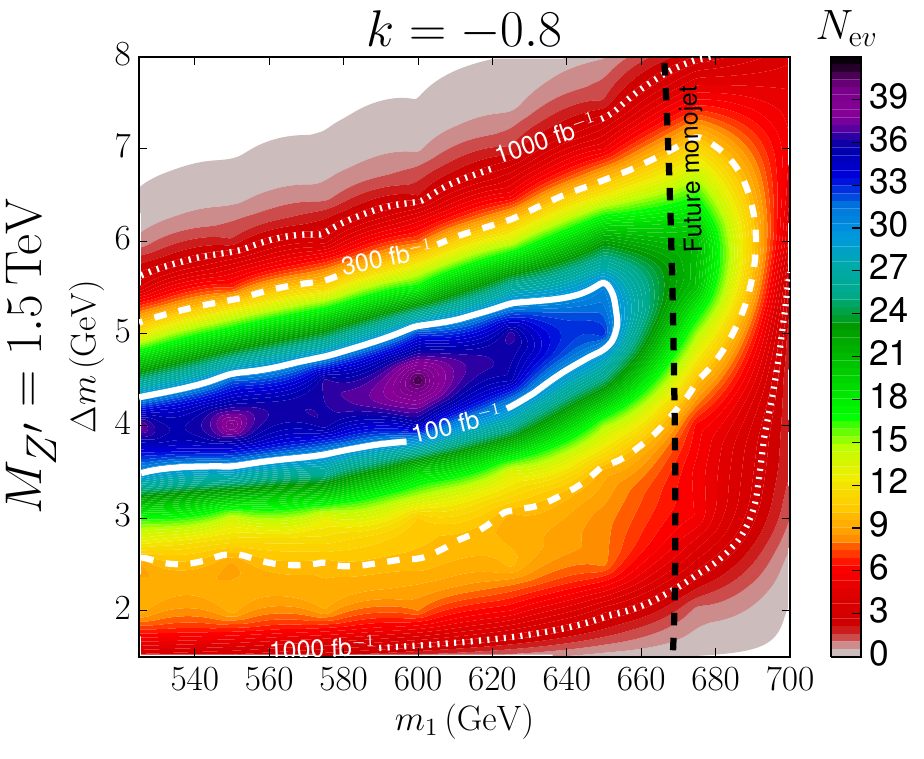}
\includegraphics[width = 0.4\linewidth]{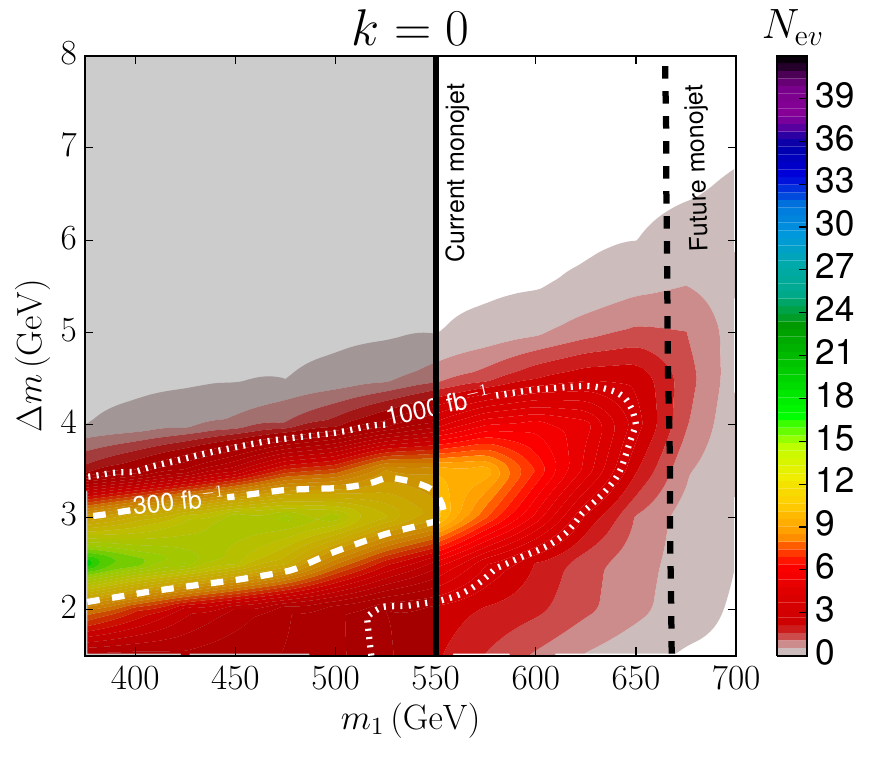}\\
\includegraphics[width = 0.425\linewidth]{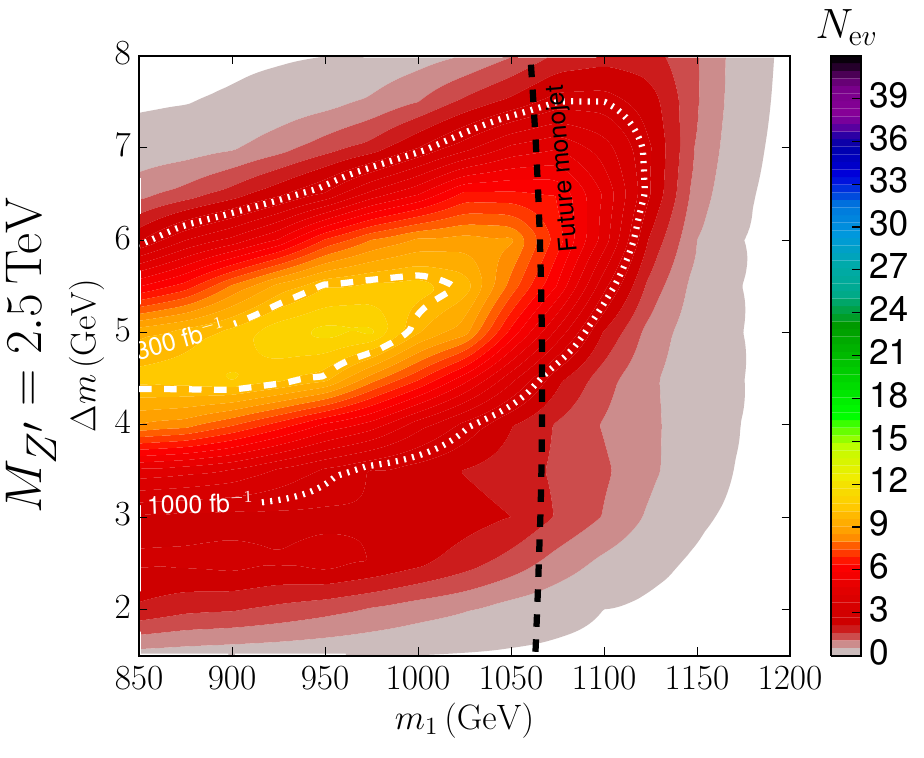}
\includegraphics[width = 0.4\linewidth]{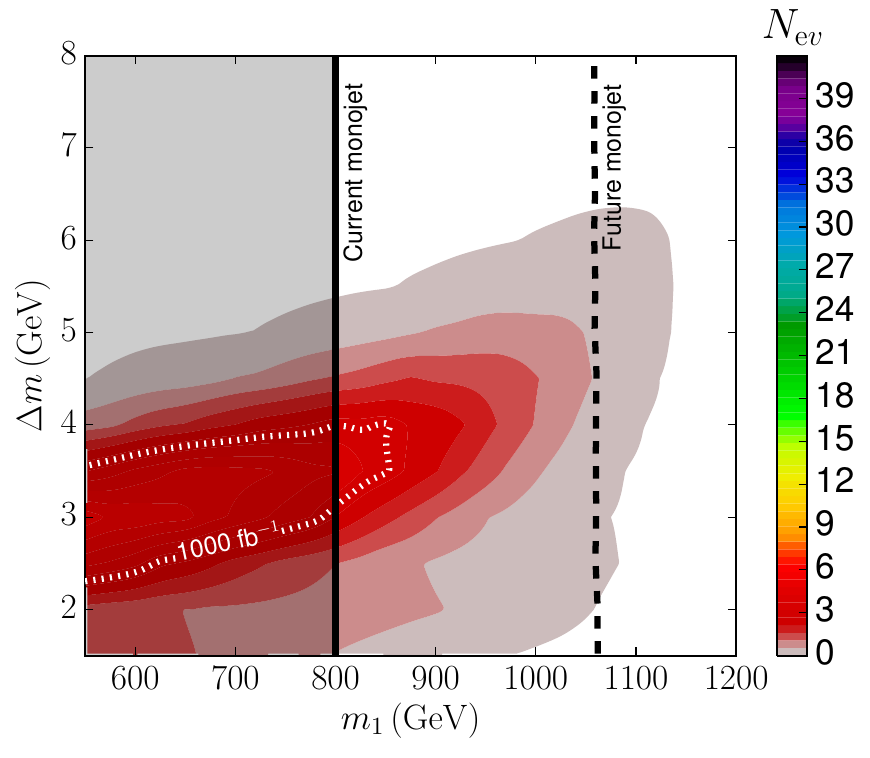}\\
\includegraphics[width = 0.425\linewidth]{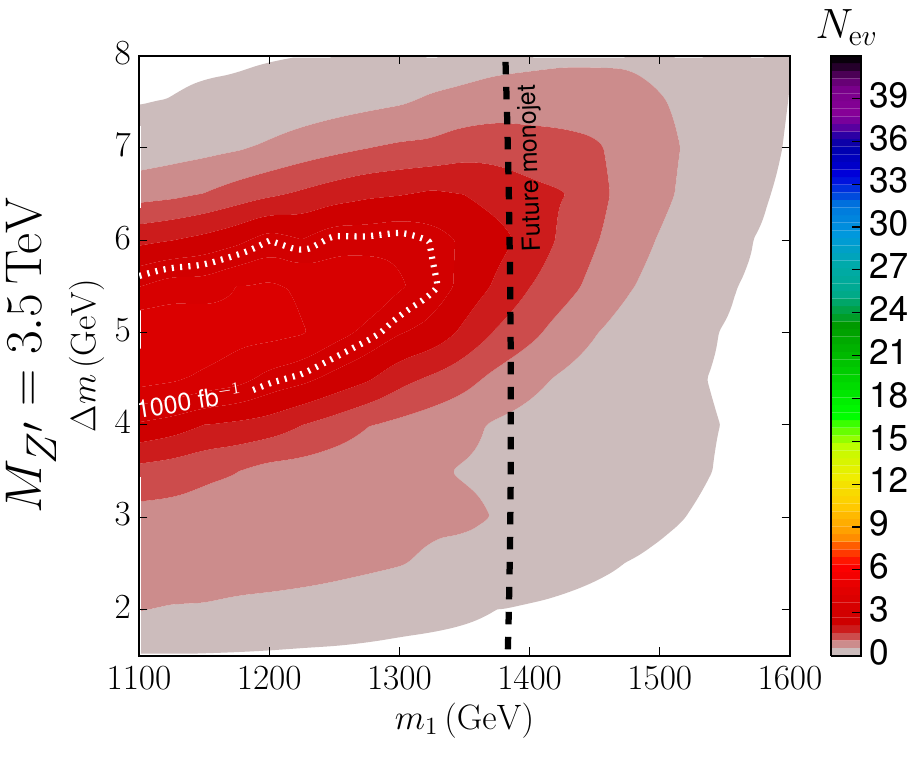}
\includegraphics[width = 0.4\linewidth]{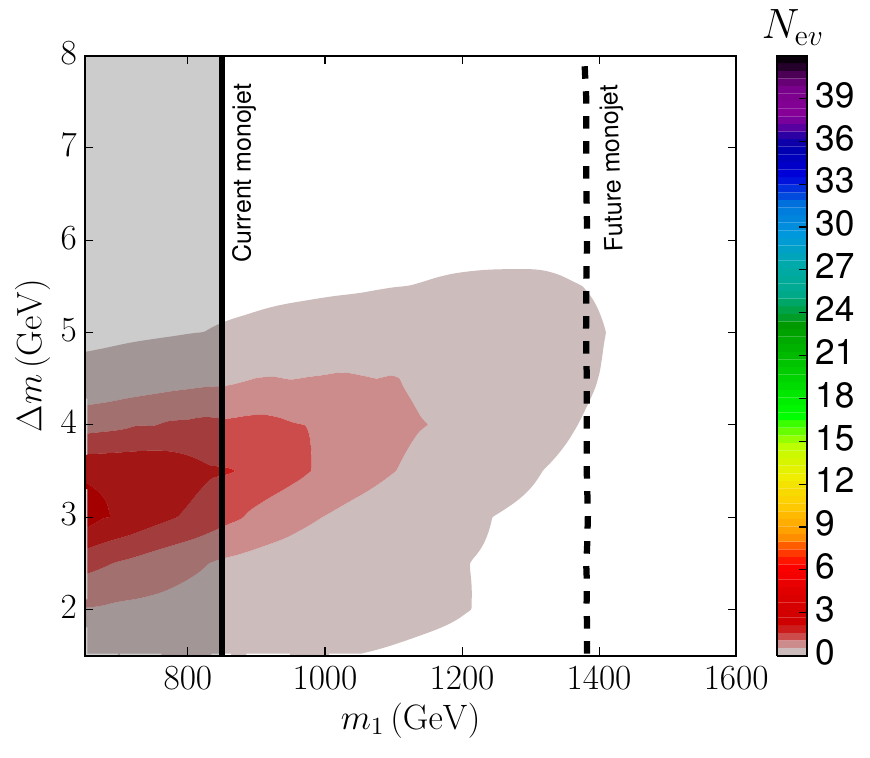}
\caption{\label{fig:dv_limits} Expected number of events and associated expected exclusion regions based on displaced vertex analysis at center of mass energy 13 TeV. The colourbar shows the expected number of events passing the displaced vertex selection criteria (see text for details) assuming $\mathcal{L}=\SI{1000}{\per\femto\barn}$. White \{dotted, dashed, solid\} lines are prospective 95\% exclusion regions at $\mathcal{L}=\{1000, 300, 100\}\,\SI{}{\per\femto\barn}$ respectively, corresponding to more than 3 events. Rows are for $M_{Z'} =\SI{1.5}{\tera\electronvolt}$ (top), $\SI{2.5}{\tera\electronvolt}$ (center) and $\SI{3.5}{\tera\electronvolt}$ (bottom). Columns are for $k = -0.8$ (left) and $k=0$ (right).  Also shown are current and prospective monojet bounds (see Section \ref{sec:monojet} for details).}
\end{figure}

In Fig.~\ref{fig:dv_limits} we apply the cuts on leading jet $p_T$ and missing energy to estimate the expected number of events for integrated luminosity $\mathcal{L}=\SI{1000}{\per\femto\barn}$ and vertex identification efficiency of $20\%$, given approximately zero expected background. In the same figure we show the potential regions of parameter space that could be ruled out at 95\% C.L. (corresponding to number of events larger than 3, with zero background) for a range of values of $\mathcal{L}$, representing a span from conservative to optimistic reach. The sensitivity of future monojet searches is shown in the figure as a dashed black line corresponding to a benchmark choice of 10\% total uncertainty in the SM backgrounds, which would correspond to limits on the model's cross-section of 5 fb. For $k=0$, the region already excluded by existing monojet searches is shaded in grey and bounded by a solid black line.

As expected, the choice of $k$ has a strong effect on the strength of the displaced vertex signal. As $k\rightarrow -1$, the decay length increases, leading to a larger number of decays within the detector volume, until at $k=-1$ the $\chi_1 \chi_2 Z'$ coupling disappears and $\chi_2$ is stable. At the same time, as $k\rightarrow -1$ the vector $\chi_1 \chi_1 Z'$ coupling increases, maximising the production cross-section and increasing the signal. Therefore the strongest constraints come when $k$ is close to -1 but not so close that the average decay length falls outside the detector.

It is interesting to see that while the signal strength is strongest for smaller values of the mediator mass $M_{Z'}$, larger values of $M_{Z'}$ allow us to also probe larger values of the DM mass $m_1$.

\section{Conclusions and outlook}\label{sec:concls}
 
In this paper, we have assessed the detectability of various signatures of pseudo-Dirac dark matter. The model is compelling as it naturally provides the correct relic density while evading direct detection constraints, at a scale which could provide striking LHC signatures.

By imposing current constraints from dijet-resonance searches for a massive $Z'$, and the relic density condition, we obtained a region of natural but as-yet-unexplored parameter space.
We studied the sensitivity of monojet and displaced vertex searches to this parameter space, finding that monojet searches are already beginning to constrain it. With greater luminosity, we expect signals or exclusions across a large mass range. 

Whilst this is attractive, the jets + missing energy signature associated with so-called `monojet' searches is a generic signal expected across a broad range of models of the dark sector. Displaced vertices are a natural companion channel, providing a smoking gun for a specific class of models. Based on our study, across the $Z'$ mass-range we consider, most of the thermal relic region of the pseudo-Dirac dark matter parameter space will first lead to a signal in the monojet channel, before eventually yielding a displaced vertex signal. 

In the event of hints of a signal in the monojet channel,
displaced vertices offer an attractive complementary search channel to characterize the features of the underlying new physics.

The model we explore could be extended to include coupling of the dark sector to leptons, which would add additional channels and constraints both from dilepton resonance searches and from displaced lepton pairs.

We stress the importance of broadening the program of DM searches at the LHC by including relatively less explored signatures such as displaced vertices, as has also been recently emphasized in Ref.~\cite{Buchmueller:2017uqu}.
The case of pseudo-Dirac DM analysed in this paper, providing the desirable features within a minimal setup, can serve as a useful benchmark model for this kind of searches.


 \section*{Acknowledgements} 
We thank  Daniele Barducci, Andrea Coccaro, Olivier Mattelaer and Iacopo Vivarelli for illuminating discussions.  The
work of VS is supported by the Science Technology and Facilities Council (STFC) under grant number ST/J000477/1. AD would like to thank Mihael Peta\v c and Jiaxin Wang for fruitful discussions.

\appendix 
\section{Full expressions for decay widths and cross sections}\label{appendix_1}
In this appendix, we provide some formulas which are used in the analysis for the displaced vertex and monojet searches.

In the limit $m_f,\Delta m\ll m_1$, we can approximate the decay width for the process $\chi_2\rightarrow\chi_1f\bar f$ as:
\begin{align}\notag
\Gamma_{\chi_2\to\chi_1\bar ff}&=\sum_f\frac{N_c^{(f)}}{480\pi^3}\,{\left(c_L+c_R\right)}^2\,\frac{\Delta m^5}{M_{Z'}^4}\Bigg\{\left(1-\frac{3}{2}\,\frac{\Delta m}{m_1}\right)\left({c_L^{(f)}}^2+{c_R^{(f)}}^2\right)\\[.15truecm]\notag
&\quad-\frac{m_f^2}{2m_1^2}\Bigg[\left(36{c_L^{(f)}}^2+33c_L^{(f)}c_R^{(f)}+36{c_R^{(f)}}^2\right)+\frac{16m_1^2}{M_{Z'}^2}\left(2{c_L^{(f)}}^2+c_L^{(f)}c_R^{(f)}+2{c_R^{(f)}}^2\right)\\[.15truecm]\notag
&\quad+\frac{10m_1^2}{\Delta m^2}\left(1-\frac{3}{2}\,\frac{\Delta m}{m_1}\right){\left({c_L^{(f)}}+{c_R^{(f)}}\right)}^2-\frac{65}{2}\,\frac{\Delta m}{m_1}\left(2{c_L^{(f)}}^2+c_L^{(f)}c_R^{(f)}+2{c_R^{(f)}}^2\right)\\[.15truecm]
&\quad-\,\frac{24m_1\Delta m}{M_{Z'}^2}\left(2{c_L^{(f)}}^2+c_L^{(f)}c_R^{(f)}+2{c_R^{(f)}}^2\right)\Bigg]\Bigg\}+\mathcal O\left[{\left(\frac{\Delta m}{m_1}\right)}^7\right]+\mathcal O\left[{\left(\frac{m_f}{m_1}\right)}^4\right]\,.
\label{decay_width_chi_2}
\end{align}
The thermally averaged cross sections for the processes $\chi_i\chi_i\rightarrow f\bar f$ and $\chi_1\chi_2\rightarrow f\bar f$ are, respectively:
\begin{subequations}
\begin{align}\notag
\langle\sigma v\rangle_{12}&=\sum_f\frac{N_c^{(f)}}{32\pi}\;\frac{{(c_L+c_R)}^2}{{\left(1-\dfrac{{(m_1+m_2)}^2}{M_{Z'}^2}\right)}^2}\;\frac{{(m_1+m_2)}^2}{M_{Z'}^4}\sqrt{1-\frac{4m_f^2}{{\left(m_1+m_2\right)^2}}}\\[.15truecm]
&\quad\Bigg[\left({c_L^{(f)}}^2+{c_R^{(f)}}^2\right)-\frac{m_f^2}{{(m_1+m_2)}^2}\left({c_L^{(f)}}^2-6c_L^{(f)}c_R^{(f)}+{c_R^{(f)}}^2\right)\Bigg]+\mathcal O\left(\frac{1}{x}\right)\label{sigma_12}\\[.2truecm]\notag
\langle\sigma v\rangle_{ii}&=\sum_f\frac{N_c^{(f)}}{8\pi}\;\frac{{(c_L-c_R)}^2}{{\left(1-\dfrac{2m_i^2}{M_{Z'}^2}\,\dfrac{2x_i+3}{x_i}\right)}^2}\,\frac{m_i^2}{M_{Z'}^4}\sqrt{1-\frac{2m_f^2}{m_1^2}\,\frac{x_i}{2x_i+3}}\\[.15truecm]
&\quad\left[\frac{{c_L^{(f)}}^2+{c_R^{(f)}}^2}{x_i}+\frac{m_f^2}{2m_1^2}{\left(c_L^{(f)}-c_R^{(f)}\right)}^2\frac{x_i}{2x_i+3}\right]\,,\label{sigma_ii}
\end{align}
\end{subequations}
where $x_1\equiv x=m_1/T$ and $x_2\equiv x\left(1+\Delta m/m_1\right)$.

In addition, in Section \ref{sec:params}, we took into account the ratio $\Gamma_{Z'}/M_{Z'}$ for the determination of the couplings between the dark sector and the SM.

Again from eqs. \eqref{pDDMLintff}-\eqref{pDDMLint11}, the partial widths for the different channels can be computed analitycally; the result is:
\begin{subequations}
\begin{align}
&\Gamma_{Z'\to\chi_1\chi_2}=\frac{(c_L+c_R)^2}{48\pi}\,M_{Z'}\,K\left[1+\frac{(m_1+m_2)^2}{2M_{Z'}^2}\right]\left(1-\dfrac{m_2-m_1}{M_{Z'}}\right)\left(1+\dfrac{m_2-m_1}{M_{Z'}}\right)\label{Z_to_12}\\[.2truecm]
&\Gamma_{Z'\to\chi_i\chi_i}=\frac{(c_R-c_L)^2}{96\pi}\,M_{Z'}\,{\left(1-\frac{4m_i^2}{M_{Z'}^2}\right)}^{\frac{3}{2}}\label{Z_to_11}\\[.2truecm]
&\Gamma_{Z'\to\bar ff}=\sum_f\frac{N_c^{(f)}}{24\pi}\,M_{Z'}\sqrt{1-\frac{4m_f^2}{M_{Z'}^2}}\left[\left({c_L^{(f)}}^2+{c_R^{(f)}}^2\right)-\frac{m_f^2}{M_{Z'}^2}\left({c_L^{(f)}}^2-6c_L^{(f)}c_R^{(f)}+{c_R^{(f)}}^2\right)\right]\label{Z_to_fermions}\,,
\end{align}
\end{subequations}
where
\begin{equation}
K\equiv\sqrt{1-2\,\frac{m_1^2+m_2^2}{M_{Z'}^2}+{\left(\frac{m_2^2-m_1^2}{M_{Z'}^2}\right)}^2}\label{K}
\end{equation}
Finally, we can compute the thermal averaged DM-fermion scattering cross-section in the non-relativistic limit, giving:
\begin{equation}
\langle\sigma v\rangle_{\chi_1f\to\chi_1f}=\sum_f\frac{N_c^{(f)}}{16\pi}{(c_L-c_R)}^2{(c_L^{(f)}-c_R^{(f)})}^2\,\frac{\mu_{\chi_1f}^2}{M_{Z'}^4}\:v\,,
\label{sigma_chi_1_f}
\end{equation}
with $\mu_{\chi_1f}=\frac{m_1m_f}{m_1+m_f}$ being the dark matter-fermion reduced mass.\\
As we can see, this scattering cross-section is both velocity and helicity suppressed, and hence it is subdominant with respect to the (co)annihilations.

\section{Details of the analysis}\label{appendix_2}
The simulations for the displaced vertex and monojet analysis are made by means of {\sc MG5\_}a{\sc MC@NLO}v2.4.2; we limit ourselves to a parton level analysis.

For the displaced vertex searches, we consider the process $pp\to\chi_2\chi_2j\to\chi_1\chi_1+5j$ via the decay $\chi_2 \to \chi_1 j j$, where $j$ generically stands for jet. As described in Section~\ref{sec:dv}, we consider this process due to the extremely low background, which occurs due to the presence of large amounts of missing energy, large jet $p_T$, and two displaced vertices. 
We handle the decay of the $\chi_2$ particle with the following steps:
\begin{enumerate}

\item we first generate 20k $pp\to\chi_2\chi_2j$ events, with $\SI{13}{\tera\electronvolt}$ c.o.m. energy. Here $j$ stands for the default multiparticle state containing the first two families quarks and the gluon;

\item we then generate 40k $\chi_2\to\chi_1jj$ events; since we consider $\SI{1.5}{\giga\electronvolt}\leq\Delta m\leq\SI{8.0}{\giga\electronvolt}$, the $b$ and $t$ quarks kinematically cannot be produced in this event;

\item we then merge these two sets of events, replacing the $\chi_2$ in the $2\to 3$ process with it's decay products, which we boost from the $\chi_2$ rest frame into the lab frame by scaling the momenta and energy by $\vec{\beta}\gamma=\vec{p}_{\chi_2}/m_{\chi_2}$ and $\gamma = E_{\chi_2}/m_{\chi_2}$ respectively. We then obtain a system of 7 particles in the final state which, for our purposes, is physically equivalent to the one we would have obtained if we had run the full process at the level of MadGraph. We have tested this procedure against direct decay of the $\chi_2$ within the full $2 \to 7$-body process, and with decay of the $\chi_2$ particle by interfacing the output $2 \to 3$-body .lhe file with {\sc BRIDGE} \cite{Meade:2007js}, finding the equivalent final kinematic distributions in all cases, with our procedure substantially faster than direct $2\to 7$ production in MadGraph\footnote{In the case of direct $2\to 7$ production in MadGraph, the extremely small width of the $\chi_2$ leads to an error in the final kinematic distributions. This is corrected by upscaling the width in the parameter card by some factor, and rescaling the final cross-section by the same factor \cite{Launchpad}.}.
\end{enumerate}

The vertex and jet identification efficiency is model-dependent and depends on the details of the detector \cite{Aad:2015uaa}, which we approximate by applying a relatively conservative flat efficiency of 20\%.

 \bibliographystyle{JHEP}
 \bibliography{bibliography} 

\providecommand{\href}[2]{#2}\begingroup\raggedright\begin{thebibliography}{10}

\bibitem{Duerr:2016tmh}
M.~Duerr, F.~Kahlhoefer, K.~Schmidt-Hoberg, T.~Schwetz and S.~Vogl, \emph{{How
  to save the WIMP: global analysis of a dark matter model with two s-channel
  mediators}}, \href{http://dx.doi.org/10.1007/JHEP09(2016)042}{\emph{JHEP}
  {\bf 09} (2016) 042}, [\href{http://arxiv.org/abs/1606.07609}{{\tt
  1606.07609}}].

\bibitem{Alves:2015pea}
A.~Alves, A.~Berlin, S.~Profumo and F.~S. Queiroz, \emph{{Dark Matter
  Complementarity and the Z$^\prime$ Portal}},
  \href{http://dx.doi.org/10.1103/PhysRevD.92.083004}{\emph{Phys. Rev.} {\bf
  D92} (2015) 083004}, [\href{http://arxiv.org/abs/1501.03490}{{\tt
  1501.03490}}].

\bibitem{Akerib:2016vxi}
{\scshape LUX} collaboration, D.~S. Akerib et~al., \emph{{Results from a search
  for dark matter in the complete LUX exposure}},
  \href{http://dx.doi.org/10.1103/PhysRevLett.118.021303}{\emph{Phys. Rev.
  Lett.} {\bf 118} (2017) 021303}, [\href{http://arxiv.org/abs/1608.07648}{{\tt
  1608.07648}}].

\bibitem{Tan:2016zwf}
{\scshape PandaX-II} collaboration, A.~Tan et~al., \emph{{Dark Matter Results
  from First 98.7 Days of Data from the PandaX-II Experiment}},
  \href{http://dx.doi.org/10.1103/PhysRevLett.117.121303}{\emph{Phys. Rev.
  Lett.} {\bf 117} (2016) 121303}, [\href{http://arxiv.org/abs/1607.07400}{{\tt
  1607.07400}}].

\bibitem{Aprile:2016swn}
{\scshape XENON100} collaboration, E.~Aprile et~al., \emph{{XENON100 Dark
  Matter Results from a Combination of 477 Live Days}},
  \href{http://dx.doi.org/10.1103/PhysRevD.94.122001}{\emph{Phys. Rev.} {\bf
  D94} (2016) 122001}, [\href{http://arxiv.org/abs/1609.06154}{{\tt
  1609.06154}}].

\bibitem{DeSimone:2010tf}
A.~De~Simone, V.~Sanz and H.~P. Sato, \emph{{Pseudo-Dirac Dark Matter Leaves a
  Trace}}, \href{http://dx.doi.org/10.1103/PhysRevLett.105.121802}{\emph{Phys.
  Rev. Lett.} {\bf 105} (2010) 121802},
  [\href{http://arxiv.org/abs/1004.1567}{{\tt 1004.1567}}].

\bibitem{Abel:2011dc}
S.~Abel and M.~Goodsell, \emph{{Easy Dirac Gauginos}},
  \href{http://dx.doi.org/10.1007/JHEP06(2011)064}{\emph{JHEP} {\bf 06} (2011)
  064}, [\href{http://arxiv.org/abs/1102.0014}{{\tt 1102.0014}}].

\bibitem{Giudice:2010wb}
G.~F. Giudice, T.~Han, K.~Wang and L.-T. Wang, \emph{{Nearly Degenerate
  Gauginos and Dark Matter at the LHC}},
  \href{http://dx.doi.org/10.1103/PhysRevD.81.115011}{\emph{Phys. Rev.} {\bf
  D81} (2010) 115011}, [\href{http://arxiv.org/abs/1004.4902}{{\tt
  1004.4902}}].

\bibitem{Fox:2002bu}
P.~J. Fox, A.~E. Nelson and N.~Weiner, \emph{{Dirac gaugino masses and
  supersoft supersymmetry breaking}},
  \href{http://dx.doi.org/10.1088/1126-6708/2002/08/035}{\emph{JHEP} {\bf 08}
  (2002) 035}, [\href{http://arxiv.org/abs/hep-ph/0206096}{{\tt
  hep-ph/0206096}}].

\bibitem{TuckerSmith:2001hy}
D.~Tucker-Smith and N.~Weiner, \emph{{Inelastic dark matter}},
  \href{http://dx.doi.org/10.1103/PhysRevD.64.043502}{\emph{Phys. Rev.} {\bf
  D64} (2001) 043502}, [\href{http://arxiv.org/abs/hep-ph/0101138}{{\tt
  hep-ph/0101138}}].

\bibitem{Nelson:2002ca}
A.~E. Nelson, N.~Rius, V.~Sanz and M.~Unsal, \emph{{The Minimal supersymmetric
  model without a mu term}},
  \href{http://dx.doi.org/10.1088/1126-6708/2002/08/039}{\emph{JHEP} {\bf 08}
  (2002) 039}, [\href{http://arxiv.org/abs/hep-ph/0206102}{{\tt
  hep-ph/0206102}}].

\bibitem{Abdallah:2015ter}
J.~Abdallah et~al., \emph{{Simplified Models for Dark Matter Searches at the
  LHC}}, \href{http://dx.doi.org/10.1016/j.dark.2015.08.001}{\emph{Phys. Dark
  Univ.} {\bf 9-10} (2015) 8--23}, [\href{http://arxiv.org/abs/1506.03116}{{\tt
  1506.03116}}].

\bibitem{DeSimone:2016fbz}
A.~De~Simone and T.~Jacques, \emph{{Simplified models vs. effective field
  theory approaches in dark matter searches}},
  \href{http://dx.doi.org/10.1140/epjc/s10052-016-4208-4}{\emph{Eur. Phys. J.}
  {\bf C76} (2016) 367}, [\href{http://arxiv.org/abs/1603.08002}{{\tt
  1603.08002}}].

\bibitem{Kahlhoefer:2017dnp}
F.~Kahlhoefer, \emph{{Review of LHC Dark Matter Searches}},
  \href{http://dx.doi.org/10.1142/S0217751X1730006X}{\emph{Int. J. Mod. Phys.}
  {\bf A32} (2017) 1730006}, [\href{http://arxiv.org/abs/1702.02430}{{\tt
  1702.02430}}].

\bibitem{Strassler:2006im}
M.~J. Strassler and K.~M. Zurek, \emph{{Echoes of a hidden valley at hadron
  colliders}},
  \href{http://dx.doi.org/10.1016/j.physletb.2007.06.055}{\emph{Phys. Lett.}
  {\bf B651} (2007) 374--379}, [\href{http://arxiv.org/abs/hep-ph/0604261}{{\tt
  hep-ph/0604261}}].

\bibitem{Strassler:2006ri}
M.~J. Strassler and K.~M. Zurek, \emph{{Discovering the Higgs through
  highly-displaced vertices}},
  \href{http://dx.doi.org/10.1016/j.physletb.2008.02.008}{\emph{Phys. Lett.}
  {\bf B661} (2008) 263--267}, [\href{http://arxiv.org/abs/hep-ph/0605193}{{\tt
  hep-ph/0605193}}].

\bibitem{Cheung:2009su}
C.~Cheung, J.~T. Ruderman, L.-T. Wang and I.~Yavin, \emph{{Lepton Jets in
  (Supersymmetric) Electroweak Processes}},
  \href{http://dx.doi.org/10.1007/JHEP04(2010)116}{\emph{JHEP} {\bf 04} (2010)
  116}, [\href{http://arxiv.org/abs/0909.0290}{{\tt 0909.0290}}].

\bibitem{Meade:2009mu}
P.~Meade, S.~Nussinov, M.~Papucci and T.~Volansky, \emph{{Searches for Long
  Lived Neutral Particles}},
  \href{http://dx.doi.org/10.1007/JHEP06(2010)029}{\emph{JHEP} {\bf 06} (2010)
  029}, [\href{http://arxiv.org/abs/0910.4160}{{\tt 0910.4160}}].

\bibitem{Feng:2010ij}
J.~L. Feng, M.~Kamionkowski and S.~K. Lee, \emph{{Light Gravitinos at Colliders
  and Implications for Cosmology}},
  \href{http://dx.doi.org/10.1103/PhysRevD.82.015012}{\emph{Phys. Rev.} {\bf
  D82} (2010) 015012}, [\href{http://arxiv.org/abs/1004.4213}{{\tt
  1004.4213}}].

\bibitem{Falkowski:2010cm}
A.~Falkowski, J.~T. Ruderman, T.~Volansky and J.~Zupan, \emph{{Hidden Higgs
  Decaying to Lepton Jets}},
  \href{http://dx.doi.org/10.1007/JHEP05(2010)077}{\emph{JHEP} {\bf 05} (2010)
  077}, [\href{http://arxiv.org/abs/1002.2952}{{\tt 1002.2952}}].

\bibitem{Meade:2010ji}
P.~Meade, M.~Reece and D.~Shih, \emph{{Long-Lived Neutralino NLSPs}},
  \href{http://dx.doi.org/10.1007/JHEP10(2010)067}{\emph{JHEP} {\bf 10} (2010)
  067}, [\href{http://arxiv.org/abs/1006.4575}{{\tt 1006.4575}}].

\bibitem{Meade:2011du}
P.~Meade, M.~Papucci and T.~Volansky, \emph{{Odd Tracks at Hadron Colliders}},
  \href{http://dx.doi.org/10.1103/PhysRevLett.109.031801}{\emph{Phys. Rev.
  Lett.} {\bf 109} (2012) 031801}, [\href{http://arxiv.org/abs/1103.3016}{{\tt
  1103.3016}}].

\bibitem{Aad:2012kw}
{\scshape ATLAS} collaboration, G.~Aad et~al., \emph{{Search for displaced
  muonic lepton jets from light Higgs boson decay in proton-proton collisions
  at $\sqrt{s}=7$ TeV with the ATLAS detector}},
  \href{http://dx.doi.org/10.1016/j.physletb.2013.02.058}{\emph{Phys. Lett.}
  {\bf B721} (2013) 32--50}, [\href{http://arxiv.org/abs/1210.0435}{{\tt
  1210.0435}}].

\bibitem{Aad:2013gva}
{\scshape ATLAS} collaboration, G.~Aad et~al., \emph{{Search for long-lived
  stopped R-hadrons decaying out-of-time with pp collisions using the ATLAS
  detector}}, \href{http://dx.doi.org/10.1103/PhysRevD.88.112003}{\emph{Phys.
  Rev.} {\bf D88} (2013) 112003}, [\href{http://arxiv.org/abs/1310.6584}{{\tt
  1310.6584}}].

\bibitem{Helo:2013esa}
J.~C. Helo, M.~Hirsch and S.~Kovalenko, \emph{{Heavy neutrino searches at the
  LHC with displaced vertices}},
  \href{http://dx.doi.org/10.1103/PhysRevD.89.073005,
  10.1103/PhysRevD.93.099902}{\emph{Phys. Rev.} {\bf D89} (2014) 073005},
  [\href{http://arxiv.org/abs/1312.2900}{{\tt 1312.2900}}].

\bibitem{Jaiswal:2013xra}
P.~Jaiswal, K.~Kopp and T.~Okui, \emph{{Higgs Production Amidst the LHC
  Detector}}, \href{http://dx.doi.org/10.1103/PhysRevD.87.115017}{\emph{Phys.
  Rev.} {\bf D87} (2013) 115017}, [\href{http://arxiv.org/abs/1303.1181}{{\tt
  1303.1181}}].

\bibitem{Aad:2014yea}
{\scshape ATLAS} collaboration, G.~Aad et~al., \emph{{Search for long-lived
  neutral particles decaying into lepton jets in proton-proton collisions at $
  \sqrt{s}=8 $ TeV with the ATLAS detector}},
  \href{http://dx.doi.org/10.1007/JHEP11(2014)088}{\emph{JHEP} {\bf 11} (2014)
  088}, [\href{http://arxiv.org/abs/1409.0746}{{\tt 1409.0746}}].

\bibitem{Buckley:2014ika}
M.~R. Buckley, V.~Halyo and P.~Lujan, \emph{{Don't Miss the Displaced Higgs at
  the LHC Again}},  \href{http://arxiv.org/abs/1405.2082}{{\tt 1405.2082}}.

\bibitem{Cui:2014twa}
Y.~Cui and B.~Shuve, \emph{{Probing Baryogenesis with Displaced Vertices at the
  LHC}}, \href{http://dx.doi.org/10.1007/JHEP02(2015)049}{\emph{JHEP} {\bf 02}
  (2015) 049}, [\href{http://arxiv.org/abs/1409.6729}{{\tt 1409.6729}}].

\bibitem{CMS:2014wda}
{\scshape CMS} collaboration, V.~Khachatryan et~al., \emph{{Search for
  Long-Lived Neutral Particles Decaying to Quark-Antiquark Pairs in
  Proton-Proton Collisions at $\sqrt{s} =$ 8 TeV}},
  \href{http://dx.doi.org/10.1103/PhysRevD.91.012007}{\emph{Phys. Rev.} {\bf
  D91} (2015) 012007}, [\href{http://arxiv.org/abs/1411.6530}{{\tt
  1411.6530}}].

\bibitem{Aad:2015rba}
{\scshape ATLAS} collaboration, G.~Aad et~al., \emph{{Search for massive,
  long-lived particles using multitrack displaced vertices or displaced lepton
  pairs in pp collisions at $\sqrt{s}$ = 8 TeV with the ATLAS detector}},
  \href{http://dx.doi.org/10.1103/PhysRevD.92.072004}{\emph{Phys. Rev.} {\bf
  D92} (2015) 072004}, [\href{http://arxiv.org/abs/1504.05162}{{\tt
  1504.05162}}].

\bibitem{Clarke:2015ala}
J.~D. Clarke, \emph{{Constraining portals with displaced Higgs decay searches
  at the LHC}}, \href{http://dx.doi.org/10.1007/JHEP10(2015)061}{\emph{JHEP}
  {\bf 10} (2015) 061}, [\href{http://arxiv.org/abs/1505.00063}{{\tt
  1505.00063}}].

\bibitem{Csaki:2015fba}
C.~Csaki, E.~Kuflik, S.~Lombardo and O.~Slone, \emph{{Searching for displaced
  Higgs boson decays}},
  \href{http://dx.doi.org/10.1103/PhysRevD.92.073008}{\emph{Phys. Rev.} {\bf
  D92} (2015) 073008}, [\href{http://arxiv.org/abs/1508.01522}{{\tt
  1508.01522}}].

\bibitem{Csaki:2015uza}
C.~Csaki, E.~Kuflik, S.~Lombardo, O.~Slone and T.~Volansky,
  \emph{{Phenomenology of a Long-Lived LSP with R-Parity Violation}},
  \href{http://dx.doi.org/10.1007/JHEP08(2015)016}{\emph{JHEP} {\bf 08} (2015)
  016}, [\href{http://arxiv.org/abs/1505.00784}{{\tt 1505.00784}}].

\bibitem{Curtin:2015fna}
D.~Curtin and C.~B. Verhaaren, \emph{{Discovering Uncolored Naturalness in
  Exotic Higgs Decays}},
  \href{http://dx.doi.org/10.1007/JHEP12(2015)072}{\emph{JHEP} {\bf 12} (2015)
  072}, [\href{http://arxiv.org/abs/1506.06141}{{\tt 1506.06141}}].

\bibitem{Khachatryan:2015jha}
{\scshape CMS} collaboration, V.~Khachatryan et~al., \emph{{Search for Decays
  of Stopped Long-Lived Particles Produced in Proton–Proton Collisions at
  $\sqrt{s}= 8\,\text {TeV} $}},
  \href{http://dx.doi.org/10.1140/epjc/s10052-015-3367-z}{\emph{Eur. Phys. J.}
  {\bf C75} (2015) 151}, [\href{http://arxiv.org/abs/1501.05603}{{\tt
  1501.05603}}].

\bibitem{Liu:2015bma}
Z.~Liu and B.~Tweedie, \emph{{The Fate of Long-Lived Superparticles with
  Hadronic Decays after LHC Run 1}},
  \href{http://dx.doi.org/10.1007/JHEP06(2015)042}{\emph{JHEP} {\bf 06} (2015)
  042}, [\href{http://arxiv.org/abs/1503.05923}{{\tt 1503.05923}}].

\bibitem{Schwaller:2015gea}
P.~Schwaller, D.~Stolarski and A.~Weiler, \emph{{Emerging Jets}},
  \href{http://dx.doi.org/10.1007/JHEP05(2015)059}{\emph{JHEP} {\bf 05} (2015)
  059}, [\href{http://arxiv.org/abs/1502.05409}{{\tt 1502.05409}}].

\bibitem{Aaboud:2016dgf}
{\scshape ATLAS} collaboration, M.~Aaboud et~al., \emph{{Search for metastable
  heavy charged particles with large ionization energy loss in pp collisions at
  $\sqrt{s} = 13$ TeV using the ATLAS experiment}},
  \href{http://dx.doi.org/10.1103/PhysRevD.93.112015}{\emph{Phys. Rev.} {\bf
  D93} (2016) 112015}, [\href{http://arxiv.org/abs/1604.04520}{{\tt
  1604.04520}}].

\bibitem{Aaboud:2016uth}
{\scshape ATLAS} collaboration, M.~Aaboud et~al., \emph{{Search for heavy
  long-lived charged $R$-hadrons with the ATLAS detector in 3.2 fb$^{-1}$ of
  proton--proton collision data at $\sqrt{s} = 13$ TeV}},
  \href{http://dx.doi.org/10.1016/j.physletb.2016.07.042}{\emph{Phys. Lett.}
  {\bf B760} (2016) 647--665}, [\href{http://arxiv.org/abs/1606.05129}{{\tt
  1606.05129}}].

\bibitem{Allanach:2016pam}
B.~C. Allanach, M.~Badziak, G.~Cottin, N.~Desai, C.~Hugonie and R.~Ziegler,
  \emph{{Prompt Signals and Displaced Vertices in Sparticle Searches for
  Next-to-Minimal Gauge Mediated Supersymmetric Models}},
  \href{http://dx.doi.org/10.1140/epjc/s10052-016-4330-3}{\emph{Eur. Phys. J.}
  {\bf C76} (2016) 482}, [\href{http://arxiv.org/abs/1606.03099}{{\tt
  1606.03099}}].

\bibitem{Coccaro:2016lnz}
A.~Coccaro, D.~Curtin, H.~J. Lubatti, H.~Russell and J.~Shelton,
  \emph{{Data-driven Model-independent Searches for Long-lived Particles at the
  LHC}}, \href{http://dx.doi.org/10.1103/PhysRevD.94.113003}{\emph{Phys. Rev.}
  {\bf D94} (2016) 113003}, [\href{http://arxiv.org/abs/1605.02742}{{\tt
  1605.02742}}].

\bibitem{Khachatryan:2016sfv}
{\scshape CMS} collaboration, V.~Khachatryan et~al., \emph{{Search for
  long-lived charged particles in proton-proton collisions at $\sqrt s=$
  13  TeV}},
  \href{http://dx.doi.org/10.1103/PhysRevD.94.112004}{\emph{Phys. Rev.} {\bf
  D94} (2016) 112004}, [\href{http://arxiv.org/abs/1609.08382}{{\tt
  1609.08382}}].

\bibitem{Mahbubani:2017gjh}
R.~Mahbubani, P.~Schwaller and J.~Zurita, \emph{{Closing the window for
  compressed Dark Sectors with disappearing charged tracks}},
  \href{http://arxiv.org/abs/1703.05327}{{\tt 1703.05327}}.

\bibitem{Buchmueller:2017uqu}
O.~Buchmueller, A.~De~Roeck, M.~McCullough, K.~Hahn, K.~Sung, P.~Schwaller
  et~al., \emph{{Simplified Models for Displaced Dark Matter Signatures}},
  \href{http://arxiv.org/abs/1704.06515}{{\tt 1704.06515}}.

\bibitem{Antusch:2017hhu}
S.~Antusch, E.~Cazzato and O.~Fischer, \emph{{Sterile neutrino searches via
  displaced vertices at LHCb}},  \href{http://arxiv.org/abs/1706.05990}{{\tt
  1706.05990}}.

\bibitem{Izaguirre:2015zva}
E.~Izaguirre, G.~Krnjaic and B.~Shuve, \emph{{Discovering Inelastic
  Thermal-Relic Dark Matter at Colliders}},
  \href{http://dx.doi.org/10.1103/PhysRevD.93.063523}{\emph{Phys. Rev.} {\bf
  D93} (2016) 063523}, [\href{http://arxiv.org/abs/1508.03050}{{\tt
  1508.03050}}].

\bibitem{Kahlhoefer:2015bea}
F.~Kahlhoefer, K.~Schmidt-Hoberg, T.~Schwetz and S.~Vogl, \emph{{Implications
  of unitarity and gauge invariance for simplified dark matter models}},
  \href{http://dx.doi.org/10.1007/JHEP02(2016)016}{\emph{JHEP} {\bf 02} (2016)
  016}, [\href{http://arxiv.org/abs/1510.02110}{{\tt 1510.02110}}].

\bibitem{Fan:2010gt}
J.~Fan, M.~Reece and L.-T. Wang, \emph{{Non-relativistic effective theory of
  dark matter direct detection}},
  \href{http://dx.doi.org/10.1088/1475-7516/2010/11/042}{\emph{JCAP} {\bf 1011}
  (2010) 042}, [\href{http://arxiv.org/abs/1008.1591}{{\tt 1008.1591}}].

\bibitem{Fitzpatrick:2012ix}
A.~L. Fitzpatrick, W.~Haxton, E.~Katz, N.~Lubbers and Y.~Xu, \emph{{The
  Effective Field Theory of Dark Matter Direct Detection}},
  \href{http://dx.doi.org/10.1088/1475-7516/2013/02/004}{\emph{JCAP} {\bf 1302}
  (2013) 004}, [\href{http://arxiv.org/abs/1203.3542}{{\tt 1203.3542}}].

\bibitem{DelNobile:2013sia}
M.~Cirelli, E.~Del~Nobile and P.~Panci, \emph{{Tools for model-independent
  bounds in direct dark matter searches}},
  \href{http://dx.doi.org/10.1088/1475-7516/2013/10/019}{\emph{JCAP} {\bf 1310}
  (2013) 019}, [\href{http://arxiv.org/abs/1307.5955}{{\tt 1307.5955}}].

\bibitem{Dent:2015zpa}
J.~B. Dent, L.~M. Krauss, J.~L. Newstead and S.~Sabharwal, \emph{{General
  analysis of direct dark matter detection: From microphysics to observational
  signatures}}, \href{http://dx.doi.org/10.1103/PhysRevD.92.063515}{\emph{Phys.
  Rev.} {\bf D92} (2015) 063515}, [\href{http://arxiv.org/abs/1505.03117}{{\tt
  1505.03117}}].

\bibitem{ElHedri:2017nny}
S.~El~Hedri, A.~Kaminska, M.~de~Vries and J.~Zurita, \emph{{Simplified
  Phenomenology for Colored Dark Sectors}},
  \href{http://arxiv.org/abs/1703.00452}{{\tt 1703.00452}}.

\bibitem{Griest:1990kh}
K.~Griest and D.~Seckel, \emph{{Three exceptions in the calculation of relic
  abundances}}, \href{http://dx.doi.org/10.1103/PhysRevD.43.3191}{\emph{Phys.
  Rev.} {\bf D43} (1991) 3191--3203}.

\bibitem{Boveia:2016mrp}
G.~Busoni et~al., \emph{{Recommendations on presenting LHC searches for missing
  transverse energy signals using simplified $s$-channel models of dark
  matter}},  \href{http://arxiv.org/abs/1603.04156}{{\tt 1603.04156}}.

\bibitem{Akerib:2017kat}
{\scshape LUX} collaboration, D.~S. Akerib et~al., \emph{{Limits on
  spin-dependent WIMP-nucleon cross section obtained from the complete LUX
  exposure}},  \href{http://arxiv.org/abs/1705.03380}{{\tt 1705.03380}}.

\bibitem{ATLAS:2016lvi}
{\scshape ATLAS} collaboration, T.~A. collaboration, \emph{{Search for New
  Phenomena in Dijet Events with the ATLAS Detector at $\sqrt{s}$=13 TeV with
  2015 and 2016 data}}, .

\bibitem{Ade:2015xua}
{\scshape Planck} collaboration, P.~A.~R. Ade et~al., \emph{{Planck 2015
  results. XIII. Cosmological parameters}},
  \href{http://arxiv.org/abs/1502.01589}{{\tt 1502.01589}}.

\bibitem{An:2012va}
H.~An, X.~Ji and L.-T. Wang, \emph{{Light Dark Matter and $Z'$ Dark Force at
  Colliders}}, \href{http://dx.doi.org/10.1007/JHEP07(2012)182}{\emph{JHEP}
  {\bf 07} (2012) 182}, [\href{http://arxiv.org/abs/1202.2894}{{\tt
  1202.2894}}].

\bibitem{Jacques:2015zha}
T.~Jacques and K.~Nordström, \emph{{Mapping monojet constraints onto
  Simplified Dark Matter Models}},
  \href{http://dx.doi.org/10.1007/JHEP06(2015)142}{\emph{JHEP} {\bf 06} (2015)
  142}, [\href{http://arxiv.org/abs/1502.05721}{{\tt 1502.05721}}].

\bibitem{Aaboud:2016tnv}
{\scshape ATLAS} collaboration, M.~Aaboud et~al., \emph{{Search for new
  phenomena in final states with an energetic jet and large missing transverse
  momentum in $pp$ collisions at $\sqrt{s}=13$  TeV using the ATLAS
  detector}}, \href{http://dx.doi.org/10.1103/PhysRevD.94.032005}{\emph{Phys.
  Rev.} {\bf D94} (2016) 032005}, [\href{http://arxiv.org/abs/1604.07773}{{\tt
  1604.07773}}].

\bibitem{Barducci:2015ffa}
D.~Barducci, A.~Belyaev, A.~K.~M. Bharucha, W.~Porod and V.~Sanz,
  \emph{{Uncovering Natural Supersymmetry via the interplay between the LHC and
  Direct Dark Matter Detection}},
  \href{http://dx.doi.org/10.1007/JHEP07(2015)066}{\emph{JHEP} {\bf 07} (2015)
  066}, [\href{http://arxiv.org/abs/1504.02472}{{\tt 1504.02472}}].

\bibitem{Aad:2015uaa}
{\scshape ATLAS} collaboration, G.~Aad et~al., \emph{{Search for long-lived,
  weakly interacting particles that decay to displaced hadronic jets in
  proton-proton collisions at $\sqrt{s}=8$ TeV with the ATLAS detector}},
  \href{http://dx.doi.org/10.1103/PhysRevD.92.012010}{\emph{Phys. Rev.} {\bf
  D92} (2015) 012010}, [\href{http://arxiv.org/abs/1504.03634}{{\tt
  1504.03634}}].

\bibitem{deFavereau:2013fsa}
{\scshape DELPHES 3} collaboration, J.~de~Favereau, C.~Delaere, P.~Demin,
  A.~Giammanco, V.~Lemaître, A.~Mertens et~al., \emph{{DELPHES 3, A modular
  framework for fast simulation of a generic collider experiment}},
  \href{http://dx.doi.org/10.1007/JHEP02(2014)057}{\emph{JHEP} {\bf 02} (2014)
  057}, [\href{http://arxiv.org/abs/1307.6346}{{\tt 1307.6346}}].

\bibitem{Meade:2007js}
P.~Meade and M.~Reece, \emph{{BRIDGE: Branching ratio inquiry / decay generated
  events}},  \href{http://arxiv.org/abs/hep-ph/0703031}{{\tt hep-ph/0703031}}.

\bibitem{Launchpad}
``Madgraph launchpad: why production and decay cross-section didn't agree.''
  \url{https://answers.launchpad.net/mg5amcnlo/+faq/2442}.

\end{thebibliography}\endgroup
 \end{document}